\documentclass[aps,pra,10pt,twocolumn,numerical]{revtex4-1}
\usepackage[utf8x]{inputenc}
\usepackage{amssymb,amsmath}
\usepackage{multirow}
\usepackage{graphicx}
\usepackage{epsfig}
\usepackage{hyperref}
\usepackage{subfigure}
\usepackage{bbm}
\usepackage{stmaryrd}
\usepackage{pspicture}
\usepackage{natbib}
\usepackage[english]{babel}
\usepackage{color}
\usepackage{soul}

\usepackage{hyperref}
\hypersetup{colorlinks=true, breaklinks=true, linkcolor=blue, urlcolor=blue}

\newcommand{\ket}[1]{\left|#1\right>}
\newcommand{\bra}[1]{\left<#1\right|}
\newcommand{\ii}{\rm{i}}

\begin{document}
\title{Dynamics of interacting transport qubits}
\author{Christian Nietner} \email{cnietner@itp.tu-berlin.de}
\author{Gernot Schaller}
\author{Christina P\"oltl}
\author{Tobias Brandes}
\affiliation{Institut f\"ur Theoretische Physik, Technische Universit\"at Berlin, Hardenbergstr. 36, 10623 Berlin, Germany}
\date{\today}
%
%
%%%%%%%%%%%%%%%%%%
%%%% ABSTRACT %%%%
%%%%%%%%%%%%%%%%%%
%
%
\begin{abstract}
 We investigate the electronic transport through two parallel double quantum dots coupled both capacitively and via a perpendicularly aligned charge qubit. The presence of the qubit leads to a modification of the coherent tunnel amplitudes of each double quantum dot. We study the influence of the qubit on the electronic steady-state currents through the system, the entanglement between the transport double quantum dots, and the back action on the charge qubit. We use a Born-Markov secular quantum master equation for the system. The obtained currents show signatures of the qubit. The stationary qubit state may be tuned and even rendered pure by applying suitable voltages. In the Coulomb diamonds it is also possible to stabilize pure entangled states of the transport double quantum dots.
\end{abstract}
\maketitle
%
%
%%%%%%%%%%%%%%%%%%%%%%%%%%%%%
\section{Introduction}   %%%%
%%%%%%%%%%%%%%%%%%%%%%%%%%%%%
%
%
The capability to harness the potential power of quantum properties, such as superposition of states and entanglement, for information and communication technologies, has raised interest in possible experimental implementations over the recent years. Scalability, the feasibility of coherent control, and non-destructive read out of quantum states as well as robustness against decoherence are among the key features for such candidates \cite{DiVincenco2000}. Encouraged by the latest progress in their fabrication and manipulation, semiconductor quantum dots (QDs) have been proposed as possible candidates \cite{Loss1998} and have been intensively studied since then.\\
Of special interest are double quantum dots (DQDs) \cite{Blick1998}, which are used to model two-level qubit states \cite{Vion2002,DiVincenco2005,Wiel2006}. It has been shown that coherent control \cite{Oosterkamp1998,Petta2005,Koppens2006,Stehlik2011} and read out \cite{Barthel2009,Petersson2010,Nowack2011} are achievable in these systems. Together, this allows us to implement electronically accessible quantum gates based on quantum dots \cite{Kohler2005,Kohler2005a,Kohler2009}. Even entanglement, which is crucial for quantum computation, can be produced \cite{Trauzettel2006}, manipulated \cite{Clive2009} and detected \cite{Brandes2007,Reuther2011} in quantum-dot setups. In consequence, quantum dots provide a promising candidate for a quantum computation architecture.\\
Nowadays, quantum dots are produced on a large scale, for example, via lithographic methods \cite{Kouwenhoven1996,Kouwenhoven1997}, self-assembled growth \cite{Klein1996}, or by depletion of two-dimensional (2D) electron gases in semiconductors \cite{Elzerman2003}, to name but a few. Unfortunately, all these fabrication methods can not completely exclude unwanted impurities that might destroy the desired properties of double quantum dots. The effects of impurities on coupled quantum-dot systems have been recently studied theoretically using molecular orbital and configuration interaction methods \cite{Nguyen2011}. Furthermore, the possibility of screening charge impurities by using multi-electron QDs were theoretically investigated \cite{Barnes2011}. In addition, spin impurities have been observed experimentally via transport spectroscopy in a carbon nanotube DQD \cite{Chorley2011}.
The main goal of this paper is to analyze the effect of impurities on the transport and system characteristics of otherwise ideal DQDs. In contrast to the previously mentioned methods, we are using an effective rate equation approach derived from a Lindblad master equation. This ansatz yields a practical means to theoretically derive the transport properties of such systems, in particular, the transport spectra. To this end, we investigate a setup of two parallel DQDs, which give rise to entangled two-particle states, and a central impurity that disturbs the intrinsic DQD tunneling amplitudes. Already without impurities, these systems may exhibit interesting
non-standard fluctuation relations \cite{Kreisbeck2010}.\\
In order to keep the calculations simple, we model the impurity as a charge qubit (CQB), which perturbs the DQD system depending of its internal state. The respective system Hamiltonian is introduced in Sec.~\ref{S:ModelHamiltonian}. Subsequently, using the master-equation approach described in Sec.~\ref{S:Liouvillian}, we analyze its influence on the electronic transport spectrum through the DQDs both numerically and theoretically within Sec.~\ref{S:Transport}. Finally, we study the purity of the system in Sec.~\ref{S:Purity} and investigate the effect of the CQB on the entanglement of the two-electron states in Sec.~\ref{S:Entanglement}.\\
%
%
%%%%%%%%%%%%%%%%%%%%%%
\section{Model}   %%%%
%%%%%%%%%%%%%%%%%%%%%%
%
%
\begin{figure}[ht]
 \includegraphics[width=0.7\columnwidth]{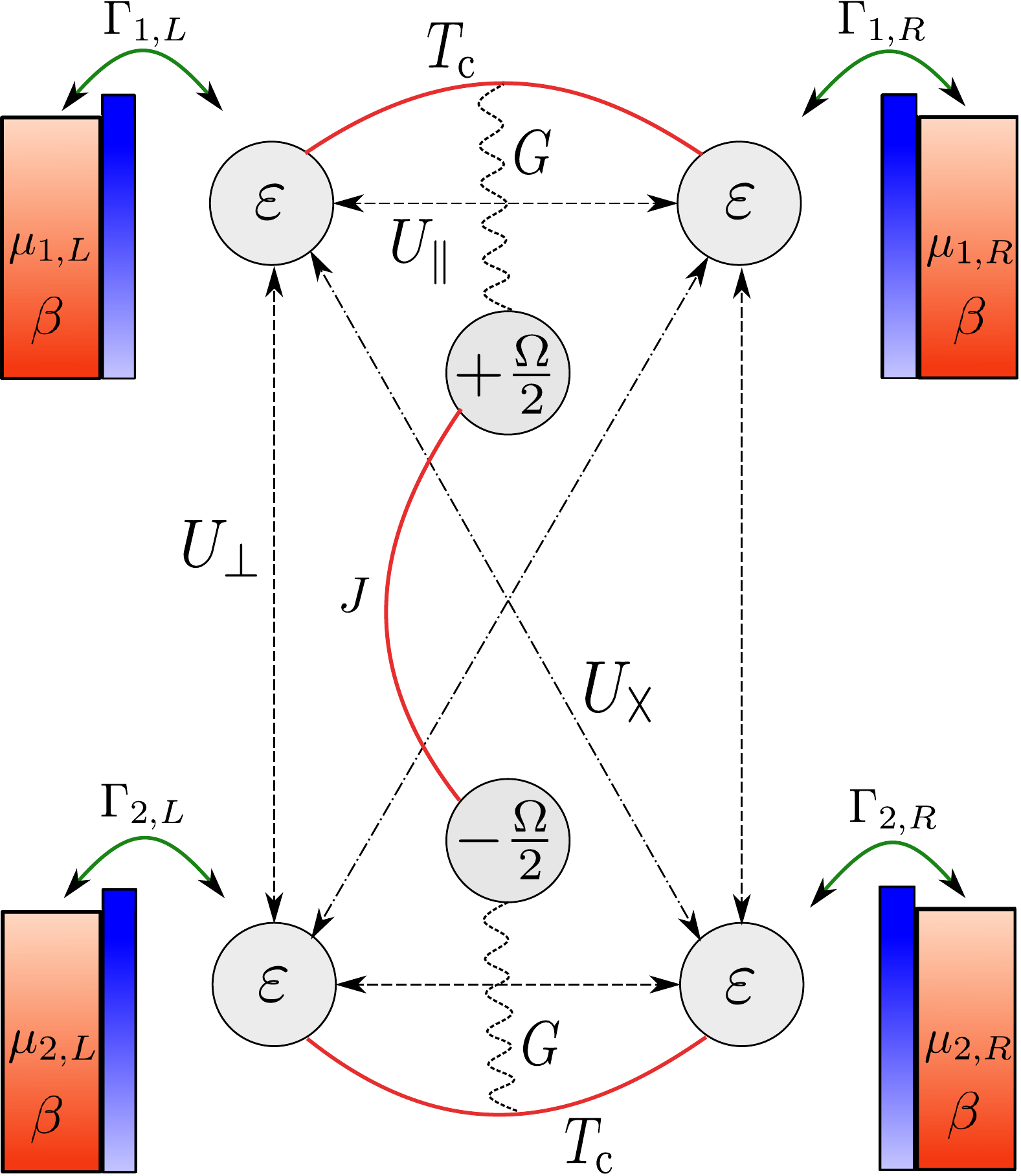}
 \caption{(Color online) Two double quantum dots (DQDs) $1$ and $2$ with electronic on-site energies $\varepsilon$, coherent tunneling amplitude (solid lines) $T_{\rm{c}}$, and on-site Coulomb interaction (dashed lines) $U_{\parallel}$ interact capacitively via perpendicular and diagonal Coulomb interactions $U_\perp$ and $U_{\vartimes}$, respectively. Transport is enabled by tunnel coupling the quantum dots $(i,\alpha)$ to adjacent fermionic reservoirs of chemical potential $\mu_{i,\alpha}$ with the tunneling rate $\Gamma_{i,\alpha}$. A charge qubit (CQB) with detuning $\Omega$ and tunneling amplitude $J$ modifies the tunnel amplitudes of both DQDs by $G$ (wavy lines) depending on the position of the CQB electron.}
 \label{F:ModelSystem}
\end{figure}
We consider an experimental setup as depicted in Fig.~\ref{F:ModelSystem}, which consists of two parallel DQDs that are coupled both capacitively and via a perpendicularly aligned CQB impurity. Furthermore, these DQDs are coupled to leads that act as reservoirs for electrons and, thus, allow for electronic transport when their parameters, i.~e.~, temperature and/or chemical potential, are chosen differently. To simplify the calculations, we explicitly assume spin-polarized electronic leads and neglect the spin degree of freedom throughout this paper. This implies that spin-selection effects as observed in Ref.~\cite{Chorley2011} do not matter.
\subsection{Hamiltonian}\label{S:ModelHamiltonian}
The Hamiltonian of the full system can be decomposed into
 $\hat{H}=\hat{H}_{\rm{S}}+\hat{H}_{\rm{B}}+\hat{H}_{\rm{SB}}$
with a system Hamiltonian $\hat{H}_{\rm{S}}$, a bath Hamiltonian $\hat{H}_{\rm{B}}$ and a Hamiltonian $\hat{H}_{\rm{SB}}$ describing the coupling between system and bath. In the considered setup, the electronic transport through each DQD is altered by the state of a nearby impurity which is modeled as a CQB. In particular, the current through the DQD closest to the charge of the CQB is suppressed due to Coulomb repulsion.\\
We define the electronic occupation number operator $\hat{n}_{\alpha}^{(i)}=\hat{d}_{i, \alpha}^{\dagger}\hat{d}_{i, \alpha}$ where the operators $\hat{d}_{i, \alpha}$ and $\hat{d}_{i, \alpha}^{\dagger}$ annihilate and create electrons in the quantum dot $\alpha \in \left\{ R,L \right\}$ in channel $i \in \left\{ 1,2 \right\}$. For convenience, we use here and in the following discussion the notation $i=1$ for the upper and $i=2$ for the lower transport channel in Fig.~\ref{F:ModelSystem}, whereas the labels $L,R$ denote the left and right quantum dots and electronic baths, respectively. Furthermore, we introduce the operators $\hat{c}_{i}$, $\hat{c}_{i}^{\dagger}$ which annihilate and create electrons in the CQB. Since the CQB is equivalent to a two-level system, we can express the corresponding operators in terms of Pauli matrices
$\hat{\sigma}_{z}=\hat{c}_{1}^{\dagger}\hat{c}_{1}-\hat{c}_{2}^{\dagger}\hat{c}_{2}$ and $\hat{\sigma}_{x}=\hat{c}_{1}^{\dagger}\hat{c}_{2}+\hat{c}_{2}^{\dagger}\hat{c}_{1}$.
Using these abbreviations yields a system Hamiltonian of the form
\begin{align}\label{E:SysHamiltonian}
 &\hat{H}_{\rm{S}}=\sum_{i=1}^2 \Big[\varepsilon \left(\hat{n}_{L}^{(i)} + \hat{n}_{R}^{(i)}\right)+T_{\rm{c}} \left( \hat{d}_{i, L}^{\dagger}\hat{d}_{i, R} + \hat{d}_{i, R}^{\dagger}\hat{d}_{i,L} \right) \notag \\
  &+U_{\parallel}\hat{n}_{L}^{(i)}\hat{n}_{R}^{(i)}\Big] \notag \\
  &+ U_\perp \left(\hat{n}_{L}^{(1)}\hat{n}_{L}^{(2)}+\hat{n}_{R}^{(1)} \hat{n}_{R}^{(2)}\right)+ U_{\vartimes} \left(\hat{n}_{L}^{(1)}\hat{n}_{R}^{(2)}+\hat{n}_{R}^{(1)} \hat{n}_{L}^{(2)}\right)\notag\\
  &+\frac{\Omega}{2}\hat{\sigma}_z + J \hat{\sigma}_x + \frac{G}{2}\Big[\left(\mathbbm{1}+\hat{\sigma}_z\right)\left(\hat{d}_{1, L}^{\dagger}\hat{d}_{1, R}+\hat{d}_{1, R}^{\dagger}\hat{d}_{1, L}\right)\notag\\
  &+ \left(\mathbbm{1}-\hat{\sigma}_z\right)\left(\hat{d}_{2, L}^{\dagger}\hat{d}_{2, R} + \hat{d}_{2, R}^{\dagger}\hat{d}_{2, L}\right)\Big].
\end{align}
The first two lines in Eq.~\eqref{E:SysHamiltonian} describe the isolated DQD properties. Specifically, the first term accounts for the homogeneous on-site energies $\varepsilon$ of the DQDs. The second term describes the intrinsic tunneling processes in the transport DQDs, which are proportional to the coherent tunneling amplitude $T_{\rm{c}}>0$. The term in the second line stems from the Coulomb interaction within each DQD that is parametrized by the Coulomb interaction strength $U_\parallel$. In the third line of Eq.~\eqref{E:SysHamiltonian}, we include the Coulomb interactions in-between the parallel DQDs that arise from electrons on the same side with strength $U_\perp$ and from electrons on different sides with strength $U_\vartimes$. Finally, the last two lines of Eq.~\eqref{E:SysHamiltonian} describe the CQB with detuning $\Omega$ and coherent tunneling amplitude $J$, and its influence on the transport DQDs:  The intrinsic processes of the DQDs are modified by the parameter $G$ in the presence of an
electron in the respective dot of the CQB. Due to Coulomb repulsion, it is reasonable to assume that the presence of an electron in the CQB leads to a suppression of the tunneling amplitude $T_{\rm{c}}$. Thus, we restrict the parameter $G$ to negative values $G\in\left[-T_{\rm{c}},0\right]$. \\
The electronic baths are modeled as free electron gases of spin-polarized electrons. Hence, the Hamiltonian for bath ($i,\alpha$) is given by
\begin{equation}\label{E:BathHamiltonian}
	\hat{H}_{\rm{B}}=\underset{i,\alpha,k}{\sum}\nu_{i,\alpha,k}\,\hat{q}_{i,\alpha,k}^{\dagger}\,\hat{q}_{i,\alpha,k},
\end{equation}
with operators $\hat{q}_{i,\alpha,k}^{\dagger}$ and $\hat{q}_{i,\alpha,k}$ creating and annihilating an electron with momentum $k$ and energy $\nu_{i,\alpha,k}$ in lead $(i,\alpha)$. The system-bath interaction Hamiltonian reads as
\begin{equation}\label{E:SysBathHamiltonian}
	\hat{H}_{\rm{SB}}=\underset{i,\alpha,k}{\sum} \left( t_{i,\alpha,k} \, \hat{q}_{i,\alpha,k}^{\dagger}\,\hat{d}_{i,\alpha}+\rm{H.\,c.} \right),
\end{equation}
where the tunneling amplitude of an electron hopping from the lead $(i,\alpha)$ into the system or vice versa is proportional to $t_{i,\alpha,k}^{*}$ and $t_{i,\alpha,k}$, respectively.
\subsection{Liouvillian}\label{S:Liouvillian}
Assuming that the leads are in thermal equilibrium and the coupling between the leads and the system is weak, we can use the Born-Markov secular approximation (BMS) \cite{Breuer}. Starting from the von Neumann equation
$\dot\rho=\mathcal{L}\left[\rho\right]$,
this formalism allows us to extract a quantum master equation that assumes the form of a rate equation for the reduced system density matrix $\rho$ in the system energy eigenbasis for non-degenerate energy eigenvalues. Within this ansatz, the Liouville super operator $\mathcal{L}$ is parametrized by the Markovian system-bath tunneling rates
$\Gamma_{i,\alpha}(\omega)=2 \pi \sum_{k} \left|t_{i,\alpha,k}\right|^2 \delta\left(\omega-\nu_{i,\alpha,k}\right)$
and the Fermi functions of the leads
%
% Definition der Fermi Funktion
$f_{i,\alpha}\left(\nu_{k}\right)=\rm{Tr_B} \big\{ \mathit{\hat{q}_{i,\alpha,k}}^{\dagger}\mathit{\hat{q}_{i,\alpha,k}}\,\rho_{B}\big\}=1/\left\{\exp\left[\beta\left(\nu_{\mathit{i,\alpha,k}} - \mu_{\alpha}\right) \right] + 1 \right\}$.
Here, we introduce the inverse temperature $\beta=1/\left(k_{\rm{B}}T\right)$, which we assume identical for each lead, and the chemical potential $\mu_{\alpha}$ for the left or right leads. For sequential electronic tunneling, we can uniquely identify the jump terms in the master equation which enables one to convert it into a conditional master equation. Considering the number $n$ of electrons tunneled via one lead ($i,\alpha$) into or out of a system with non-degenerate energy eigenvalues, the conditional master equation reads as
\begin{align}\label{E:BMSrateEquation}
 \dot\rho^{(n)}=\mathcal{L}_{0}\,\rho^{(n)}+\mathcal{L}^{+}\,\rho^{(n-1)}+\mathcal{L}^{-}\,\rho^{(n+1)}.
\end{align}
Here, the super-operators $\mathcal{L}_{0}$, $\mathcal{L}^{+}$, and $\mathcal{L}^{-}$ are acting on the reduced system density matrix. The operators $\mathcal{L}^{+}$/$\mathcal{L}^{-}$ describe electronic jumps into/from the monitored reservoir from/into the system, respectively. The operator $\mathcal{L}_{0}$ describes the internal dynamics of the system and electronic jumps between the system and the remaining baths. This $n$-resolved master equation can also be established using virtual detectors as bookkeeping operators \cite{Gernot2009}.\\
Subsequently, we perform a Fourier transformation $\rho\left(\chi_{\alpha},t\right)=\sum_n \rho^{(n)}(t) \exp(\ii \,\mathit{n}\, \chi_{\alpha})$, which introduces a counting field $\chi_{\alpha}$ for the respective lead $\alpha$. Hence, the Liouville super operator for lead $\alpha$ becomes a function of this counting field:
\begin{equation}\label{E:LiouvillianChi}
  \mathcal{L}\left(\chi_{\alpha}\right)=\mathcal{L}_{0}+\mathcal{L}^{+}e^{+\ii \chi_{\alpha}}+\mathcal{L}^{-}e^{-\ii \chi_{\alpha}}.
\end{equation}
This procedure can be applied for each of the four electronic baths leading to a corresponding number of counting fields. However, due to charge conservation and the neglect of short-time dynamics, we just need to consider one counting field for each transport channel. Hence, without loss of generality, we consider the leads $(1,L)$ and $(2,L)$ and denote the corresponding counting fields as $\chi_1$ and $\chi_2$. In the following calculations, we focus on the steady-state currents through the system. For lead $\alpha$, this current \cite{Gernot2009} is defined by the relation
\begin{equation}\label{E:StaticCurrentDefinition}
 I_{\alpha}=-\ii\,\rm{Tr}\left\lbrace\left. \partial_{\chi_{\alpha}}\mathcal{L}(\chi_{1},\chi_{2})\right|_{\chi_{1}=\chi_{2}=0}\,\bar\rho \right\rbrace,
\end{equation}
where the steady-state reduced density matrix $\bar\rho$ is determined by $0=\mathcal{L} (0,0)\,\bar\rho$. Finally, we note that throughout this paper, we utilize the flat-band limit with energy-independent tunneling rates $\Gamma_{i,\alpha}\left(\omega\right) =\Gamma_{i,\alpha}$.
%
%
%%%%%%%%%%%%%%%%%%%%%%%%%%%%%%%%%%%%%%%%%%%%%%%%%%%%%%%%%%%%%%
\section{Transport Characteristics}\label{S:Transport}    %%%%
%%%%%%%%%%%%%%%%%%%%%%%%%%%%%%%%%%%%%%%%%%%%%%%%%%%%%%%%%%%%%%
%
%
Within the following sections, we will apply the master equation formalism derived above to the Hamiltonian~\eqref{E:SysHamiltonian} and analyze its steady-state properties both analytically and numerically.
%
%
%%%%%%%%%%%%%%%%%%%%%%%%%%%%%%%%%%%%%%%%%%%%%%%%%%%
\subsection{Full Transport Characteristics}    %%%%
%%%%%%%%%%%%%%%%%%%%%%%%%%%%%%%%%%%%%%%%%%%%%%%%%%%
%
%
First, we analyze the steady-state transport spectrum of the full Hamiltonian \eqref{E:SysHamiltonian}. To this end, we evaluate Eq.~\eqref{E:StaticCurrentDefinition} numerically. This requires to set up the full $32\times32$ Liouvillian (see Appendix~\ref{S:Appendix}). The transport spectrum is obtained by calculating the steady-state current for varying gate voltage $V_{\rm{Gate}}=\varepsilon$ which shifts the energy levels of all transport DQDs according to Eq.~\eqref{E:SysHamiltonian}, and varying external bias voltage $V_{\rm{Bias}}$. The bias voltage enters the Liouville super operator through the Fermi functions via the chemical potentials. For convenience, we assume symmetric chemical potentials $\mu_{\rm L}=V_{\rm{Bias}}/2$ and $\mu_{\rm R}=-V_{\rm{Bias}}/2$ for both transport channels  in all further calculation.
\begin{figure}[ht]
   \centering \includegraphics[width=1\columnwidth]{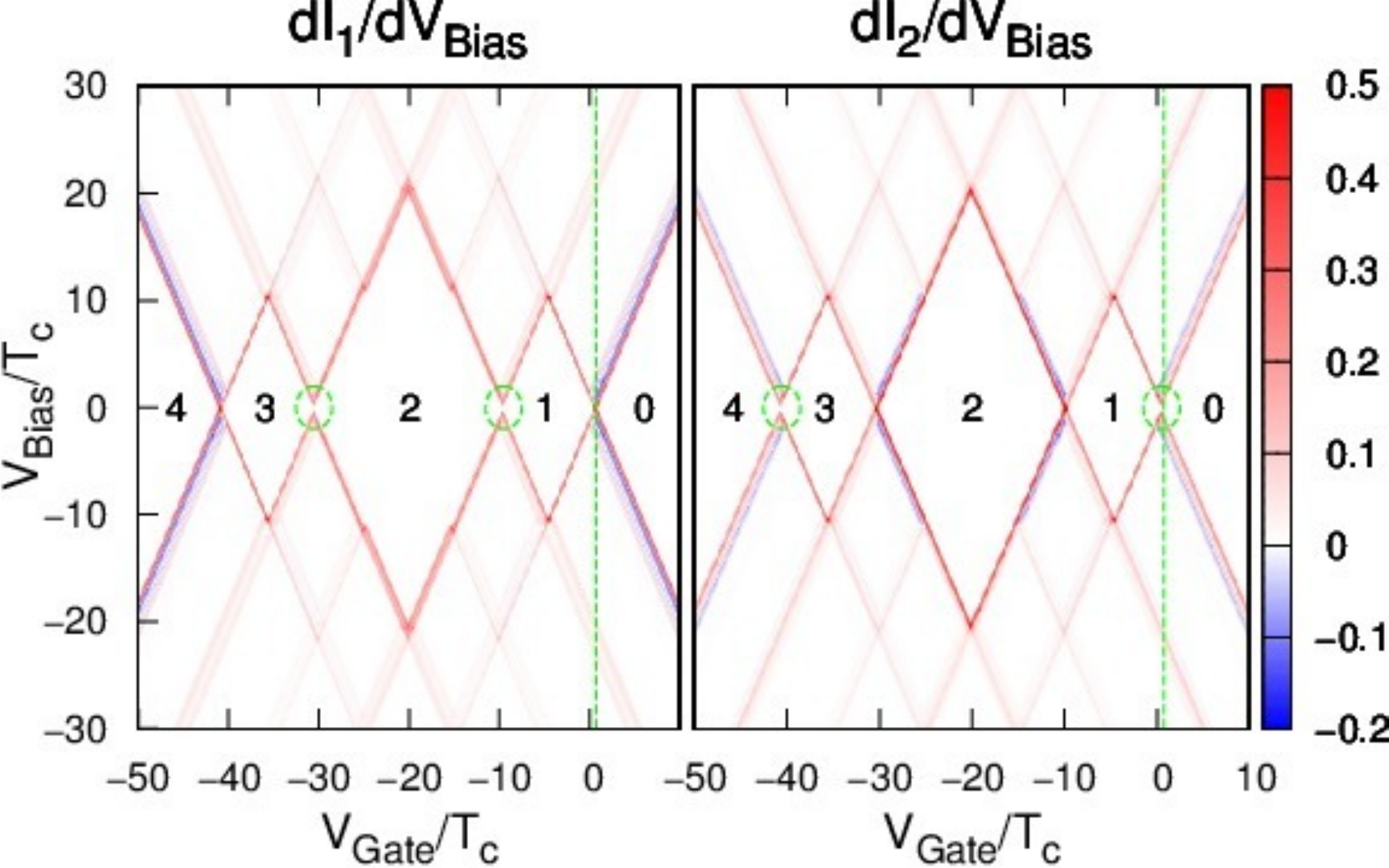}
   \caption{(Color online) Unitless differential steady-state currents through channel $1$ (left) and channel $2$ (right).  Most prominent are the CD structures with a different number of electrons (numbers) loaded into the system. The dashed circles indicate the presence of gaps between the CDs and the dashed vertical lines correspond to Fig.~\ref{F:CurrentDataPlot}. The other parameters are fixed to: $\beta=100/T_{\rm{c}}$, $U_\parallel=20\,T_{\rm{c}}$, $U_\perp=U_\vartimes=10\,T_{\rm{c}}$, $J=\Omega=\Gamma_{i,\alpha}=T_{\rm{c}}$ and $G=-T_{\rm{c}}$.} \label{F:FullTransportSpektrum}
\end{figure}
As an example, we show the differential transport spectra for both transport channels in Fig.~\ref{F:FullTransportSpektrum}. The largest differential currents $dI/dV_{\rm{Bias}}$ are observed at the edges of the  Coulomb diamonds (CDs). Within these regions, the steady-state current vanishes due to Coulomb blockade effects that prevent electronic transport through the system \cite{Livermore1996}. For a suitable choice of gate and bias voltages, this blockade can be overcome leading to finite steady-state currents. The Coulomb diamonds can be associated with specific electronic subspaces reaching from the vacuum state with zero electrons up to the maximal number of four electrons, which are indicated in Fig.~\ref{F:FullTransportSpektrum}. Note that distinct CDs can only be observed for low temperatures.
%
%
%%%%%%%%%%%%%%%%%%%%%%%%%%%%%%%%%%%%%%%
\subsection{High-Bias Currents}    %%%%
%%%%%%%%%%%%%%%%%%%%%%%%%%%%%%%%%%%%%%%
%
%
Unfortunately, the Hamiltonian \eqref{E:SysHamiltonian} is complicated to analyze analytically. Thus, we first consider the case of a high bias voltage $V_{\rm{bias}}\gg 0$, where only the system transition energies from the zero-to-one electron subspace lie within the transport window. In this ultra-strong Coulomb blockade (USCB) regime, we make use of Eq.~\eqref{E:StaticCurrentDefinition} and find that the high-bias steady-state current through the transport channel $i$ becomes
\begin{equation}
   I_{\rm{USCB}}^{(i)}=\frac{\Gamma _{i,L} \Gamma _{i,R} \Gamma _{\bar i,R}}{2\left(\Gamma _{i,L} \Gamma _{\bar i,R}+\Gamma _{i,R} \Gamma_{\bar i,L}\right)+\Gamma_{i,R} \Gamma_{\bar i,R} }\label{E:SSC1e1},
\end{equation}
where $\bar i$ labels the opposite transport channel.\\
For comparison, we additionally calculate the high-bias steady-state current for the strong Coulomb blockade (SCB) regime where in \textit{each} DQD at most one electron is allowed. In this high-bias limit, where all transition energies from the zero-to-one and one-to-two electron subspace lie within the transport window, the steady-state current through transport channel $i$ is of the form
\begin{equation}
   I_{\rm{SCB}}^{(i)}=\frac{\Gamma _{i,L} \Gamma _{i,R}}{2\, \Gamma _{i,L}+\Gamma _{i,R}}\label{E:SSC2e}.
\end{equation}
This result corresponds to the high-bias steady-state current one obtains for sequential electronic transport through a two-level system \cite{Gernot2009} in the SCB regime. From Eq.~\eqref{E:SSC1e1}, we see that in the USCB regime the current $I_{\rm{USCB}}^{(i)}$ explicitly depends on the tunneling rates of the opposite channel $\bar{i}$. This is to be expected for this configuration with dynamical channel blockade \cite{Cottet2004,Gernot2010}. This intermediate coupling is lifted, for example, if the DQD $i$ is almost immediately reloaded from the left lead, i.~e.~, $\Gamma_{i,L}\gg\Gamma_{i,R}$. Then, the system tends to be always occupied by an electron: The steady-state current in channel $i$ becomes proportional to the coupling to the right lead $\Gamma_{i,R}$, whereas the current through the other transport channel vanishes. In the opposite case, when the DQD $i$ is not refilled from the left lead, i.~e.~, $\Gamma_{i,L}\rightarrow 0$, while the other tunneling rates
remain non-vanishing, the steady-state current through channel $i$ vanishes and the other current takes on the form of Eq.~\eqref{E:SSC2e}. Considering a similar configuration where the electrons almost instantly leave the transport channel $i$ via the right lead,  i.~e.~, $\Gamma_{i,R}\gg\Gamma_{i,L}$, we find that the channel coupling is only partially lifted. In fact, in this limit, channel $\bar{i}$ decouples and the respective steady-state current takes on the form of Eq.~\eqref{E:SSC2e}. However, the steady-state current $i$ is still proportional to tunnel couplings of both transport channels. Finally, if the couplings to the left leads and the couplings to the right leads are the same for both channels, i.~e.~, $\Gamma_{i,\alpha}\rightarrow \Gamma_{\alpha}$, the steady-state currents are the same for each channel.\\
This demonstrates that in the high-bias regime, the steady-state currents are not sensitive to the asymmetry induced by the CQB. Note that due to electron-hole symmetry, we find analogous results to Eq.~\eqref{E:SSC1e1} if the system transition energies from the three-to-four electron subspace lie within the transport window.
%
%
%%%%%%%%%%%%%%%%%%%%%%%%%%%%%%%%%%%%%%%%%%%%%
\subsection{Current Anti-Correlation}    %%%%
\begin{figure}[t!]
   \centering \includegraphics[width=0.45\textwidth]{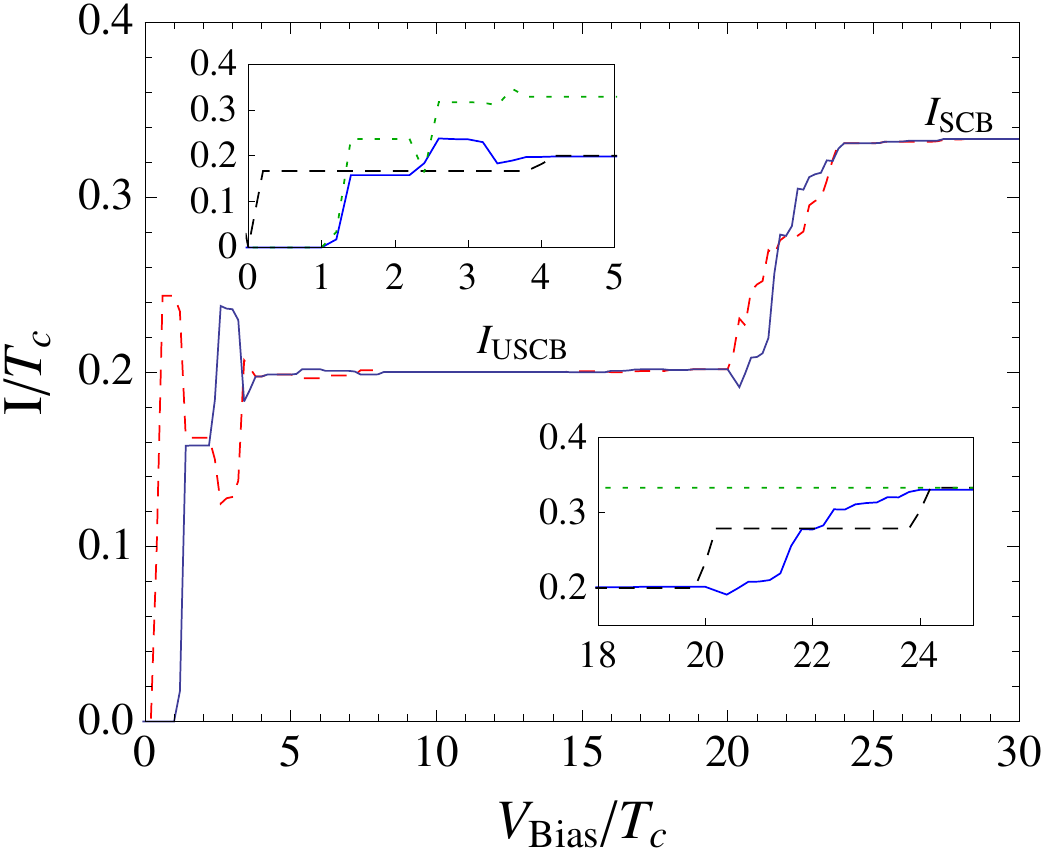}
   \caption{(Color online) Plot of the steady-state currents for both transport channels $1$ (dashed) and $2$ (solid). The plot corresponds to the dashed vertical line in Fig.~\ref{F:FullTransportSpektrum} at $V_{\rm{Gate}}=0.8\,T_{\rm{c}}$. We indicated the high-bias currents in the ultra-strong Coulomb blockade ($I_{\rm{USCB}}$) and strong Coulomb blockade ($I_{\rm{SCB}}$) regimes. The insets compare the steady-state current of channel 2 (solid) with the results for parallel DQDs without impurity (dashed) and a single DQD coupled to a CQB (dotted). The system parameters are fixed to:  $\beta=100/T_{\rm{c}}$, $U_\parallel=20\,T_{\rm{c}}$, $U_\perp=U_\vartimes=10\,T_{\rm{c}}$, $J=\Omega=\Gamma_{i,\alpha}=T_{\rm{c}}$ and $G=-T_{\rm{c}}$.}
   \label{F:CurrentDataPlot}
\end{figure}

%%%%%%%%%%%%%%%%%%%%%%%%%%%%%%%%%%%%%%%%%%%%%
%
%
In this section, we further analyze the transport properties of the system in the ultra-strong Coulomb blockade regime for small bias voltages. Thus, we focus on the region in the vicinity of the edge of the $0$-electron CD of the transport spectra in Fig.~\ref{F:FullTransportSpektrum}. A first interesting feature in this region is the occurrence of negative differential conductance, which indicates blocking effects in both transport channels. We demonstrate in the inset in Fig.~\ref{F:CurrentDataPlot} that this is an intrinsic feature resulting from the CQB impurity as negative differential conductance is not present in parallel DQDs without impurity. Effectively, the asymmetry induced by the impurity leads to the fact that the two possible transport channels become accessible at different gate voltages.
Furthermore, since the total number of electrons in the system is constrained by Coulomb interactions, as soon as the second transport channel becomes available, the current in the first channel decreases, leading to a negative differential conductance. Consistently, this property is most prominent in the Coulomb blockade regime. Therefore, by comparing the differential currents for the two transport channels, we find that this feature is accompanied by an anti-correlation between the steady-state currents in the different transport channels.
%
%
%%%%%%%%%%%%%%%%%%%%%%%%%%%%%%%%%%%%%%%%%
\subsection{Coulomb Diamond gaps}    %%%%
%%%%%%%%%%%%%%%%%%%%%%%%%%%%%%%%%%%%%%%%%
%
%
The most striking difference between the plots in Fig.~\ref{F:FullTransportSpektrum} is the fact that some Coulomb diamonds do not close, as indicated by the dashed circles. We expect this effect to occur for coupled DQDs with asymmetries in the quantum-dot energies or in their tunneling amplitudes. Because we explicitly exclude this kind of asymmetry in the transport DQDs, this effect clearly suggests a blocking induced by a finite energy barrier at zero bias voltage that stems from the presence of the CQB. In which channel the gap appears depends on the sign of the detuning.\\
Since all changes in the steady-state currents are associated with resonances of the transition energies between eigenstates of the system Hamiltonian, it is possible to calculate the position of the lines shown in Fig.~\ref{F:CurrentDataPlot} if the eigenvalues of the respective Hamiltonian $\hat{H}_{\rm S}$ are known. Hence, a diagonalization of Eq.~\eqref{E:SysHamiltonian} allows one to calculate the position and width of the gap. For example, the gate voltage with minimum gap between the $0$-electron and $1$-electron CD is given by
\begin{align}\label{E:PositionOfGap}
   V_{\rm{Gate}}^{\rm{min}}=&\frac{1}{4} \left(2\,G+4\,T_{\rm{c}}+\sqrt{4\,J^2+(G-\Omega)^2}\right.\notag\\
   &\left.-2\,\sqrt{4\,J^2+\Omega ^2}+\sqrt{4\,J^2+(G+\Omega)^2}\right).
\end{align}
Subsequently, we derive the other parameters of the gap such as the upper bias voltage at this point, which reads as
\begin{equation}
    V_{\rm{Bias}}^{\rm{upper}}=\sqrt{J^2+\left(\frac{G-\Omega}{2}\right)^2}-\sqrt{J^2+\left(\frac{G+\Omega}{2}\right)^2}.
\end{equation}
Due to the choice of our parameters, the gap is symmetric with respect to $V_{\rm{Bias}}=0$ and thus the lower bias voltage satisfies $V_{\rm{Bias}}^{\rm{lower}}=- V_{\rm{Bias}}^{\rm{upper}}$, which yields for the width of the gap
\begin{equation}\label{E:WidthOfGap}
   \Delta_{\rm{Gap}}=\sqrt{4\,J^2+(G-\Omega )^2}-\sqrt{4\,J^2+(G+\Omega )^2}.
\end{equation}
In an analogous way, we can determine the point of contact of these Coulomb diamonds. We find that this position is given by the relation
\begin{equation}\label{E:CloseOfGap}
   V_{\rm{Gate}}^{\rm{cross}}=\frac{G}{2}+T_{\rm{c}}-\sqrt{J^2+\left(\frac{\Omega}{2}\right)^2}+\sqrt{J^2+\left(\frac{G-\Omega}{2}\right)^2}.
\end{equation}
From Eq.~\eqref{E:WidthOfGap}, we see that for the considered experimental setup, the appearance of the gap results from both, the detuning $\Omega$ of the CQB and the modification $G$ of the intrinsic tunnel amplitudes. If one of these quantities is zero, the gap vanishes. On the contrary, the width of the gap is completely independent of the intrinsic tunnel amplitudes $T_{\rm{c}}$ of the transport DQDs. Moreover, we find that the gap not only vanishes as $G$ or $\Omega$ approaches zero, but also if $J$ becomes very large. This behavior can be understood since the rapidly oscillating CQB on average affects both transport channels in the same way and the energy barrier vanishes.
Investigating the position of the gap according to Eq.~\eqref{E:PositionOfGap} yields a more complicated behavior in dependence of the CQB parameter $J$. As the hopping amplitude $J$ is increased from zero, the gap is shifted from an initial finite value to higher gate voltages up to a maximal value. A further increase of $J$ results in a shift of the minimum gap to lower gate voltages, which in the limit $J\rightarrow \infty $ becomes a constant that equals the position for $J=0$. Therefore, we conclude that the system is most sensitive to the CQB for a small tunneling amplitude $J$.\\
In general, we note that measuring in an experiment the quantities described by Eqs.~\eqref{E:PositionOfGap}--\eqref{E:CloseOfGap} allows us to calculate the tunneling amplitude $J$ and the detuning $\Omega$ of the CQB as well as the modification $G$ of the tunnel amplitudes $T_{\rm{c}}$ of the DQDs.
%
%
%%%%%%%%%%%%%%%%%%%%%%%%%%%%%%%%%%%%%%%%%%%%%%%%%%%%%%%%%%%%
\section{Preparation of Pure States}\label{S:Purity}    %%%%
%%%%%%%%%%%%%%%%%%%%%%%%%%%%%%%%%%%%%%%%%%%%%%%%%%%%%%%%%%%%
%
\begin{figure}[t!]
   \centering \includegraphics[width=1\columnwidth]{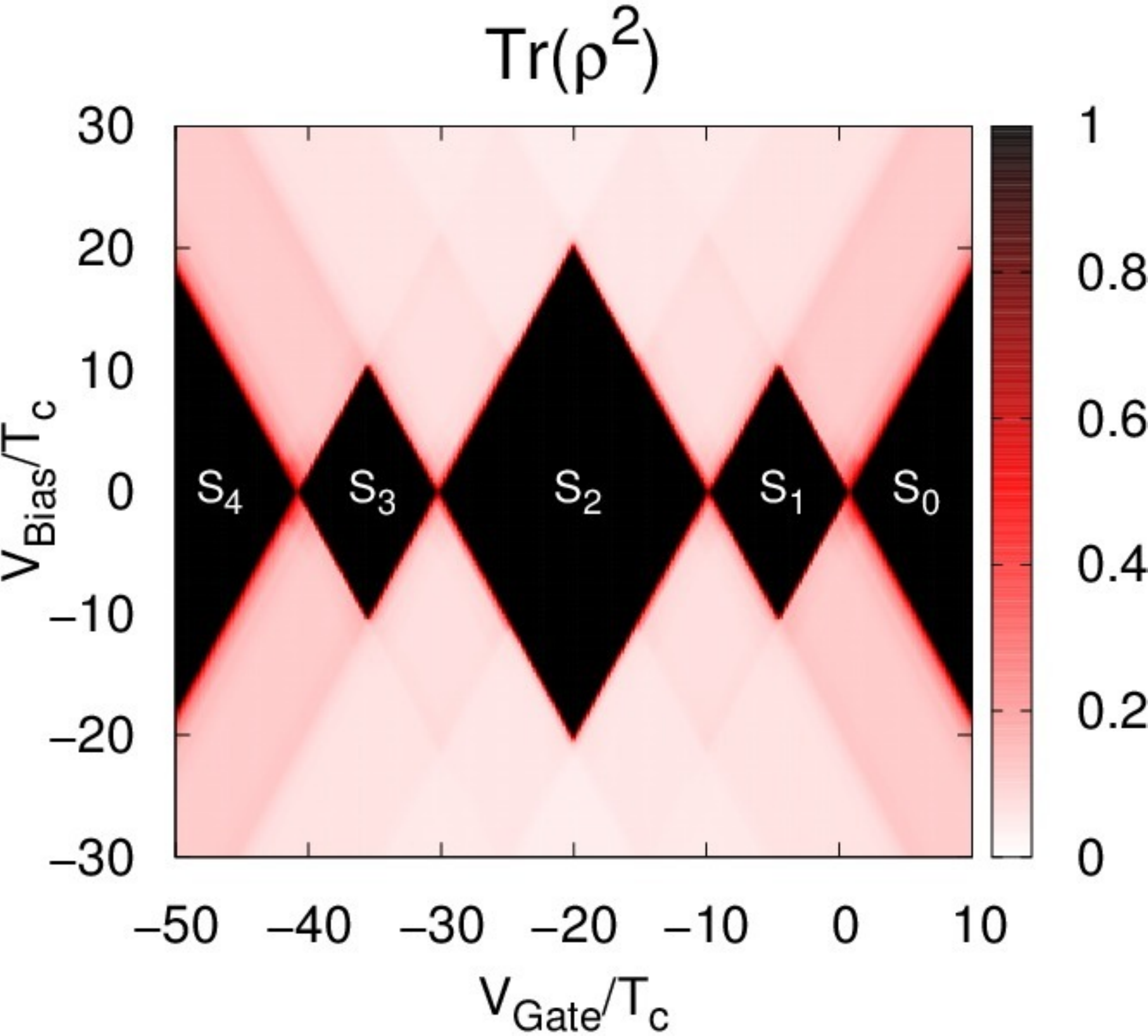}
   \caption{(Color online) Density plot of the purity of the system density matrix for low temperatures $\beta=100/T_{\rm{c}}$. The picture also shows the respective normalized eigenvectors of $\hat{H}_{\rm{S}}$ indicated in the CDs by $S_0$ ,$S_1$, $S_2$, $S_3$, and $S_4$, which correspond to Eq.~\eqref{E:Eigenstates}.
The system parameters are fixed to $U_\parallel=20\,T_{\rm{c}}$, $U_\perp=U_\vartimes=10\,T_{\rm{c}}$, $J=\Omega=\Gamma_{i,\alpha}=T_{\rm{c}}$ and $G=-T_{\rm{c}}$.}\label{F:FullPurityAnalysis}
\end{figure}
The SCB regime is particularly interesting because it includes configurations where the transport DQDs can also be treated as CQBs, giving rise to an effective system composed of three coupled qubits. These coupled qubits represent possible candidates for basic constituents of quantum information technologies that use entangled qubit states for calculations or communication. In order to better understand the eigenstates of the considered system in the long-time limit, we analyze the purity of its reduced steady-state system density matrix and investigate some interesting expectation values.\\
\subsection{Purity of the Full System}
In Fig.~\ref{F:FullPurityAnalysis}, we plot the purity $\rm{Tr}\left\{\rho^2\right\}$ of the reduced system density matrix. We find that in the interior of the Coulomb diamonds and for low temperatures, the system enters a pure energy eigenstate with the lowest possible energy. This behavior causes the vanishing steady-state currents in this region. In general, these eigenstates, which are obtained from a diagonalization (see Appendix \ref{S:Appendix}) of the Hamiltonian \eqref{E:SysHamiltonian}, are entangled states. However, for the special choice of Coulomb interactions $U_\perp=U_\vartimes$, we find that the eigenstates become separable. Therefore, the eigenstates $S_i$ the system enters in the CDs of Fig.~\ref{F:FullPurityAnalysis} have the simple representation
\begin{align}\label{E:Eigenstates}
   S_{0}=&\ket{0,0}\otimes\left(a_{0}\ket{\downarrow}-b_{0}\ket{\uparrow}\right),\notag\\
   S_{1}=&\left(\ket{L,0}-\ket{R,0}\right)\otimes\left(a_{1}\ket{\downarrow}-b_{1}\ket{\downarrow}\right),\notag\\
   S_{2}=&\left(\ket{L} - \ket{R}\right) \otimes\left(\ket{L} - \ket{R}\right)\otimes \left(a_{2}\ket{\uparrow}-b_{2}\ket{\downarrow}\right),\notag\\
   S_{3}=&\left(\ket{L,L R}-\ket{R,L R}\right)\otimes \left(a_{3}\ket{\downarrow}-b_{3}\ket{\uparrow}\right),\notag\\
   S_{4}=&\ket{LR,LR}\otimes\left(a_{4}\ket{\downarrow}-b_{4}\ket{\uparrow}\right),
\end{align}
where the structure of the kets is defined in the local basis as $\ket{\rm{DQD}_1,\rm{DQD}_2,\rm{CQB}}$ and $\left\{a_i , b_i\right\} \in \mathbbm{R}$ are the respective normalized coefficients.\\
If the temperature is raised, the region within a CD where purity is reached shrinks in favor of a mixture of states with the same number of electrons. Outside of the Coulomb diamonds, eigenstates belonging to different electron subspaces mix together, which allows for electronic transport and a finite current.
\subsection{Purity of the CQB}
\begin{figure}[t]
	\centering \includegraphics[width=1\columnwidth]{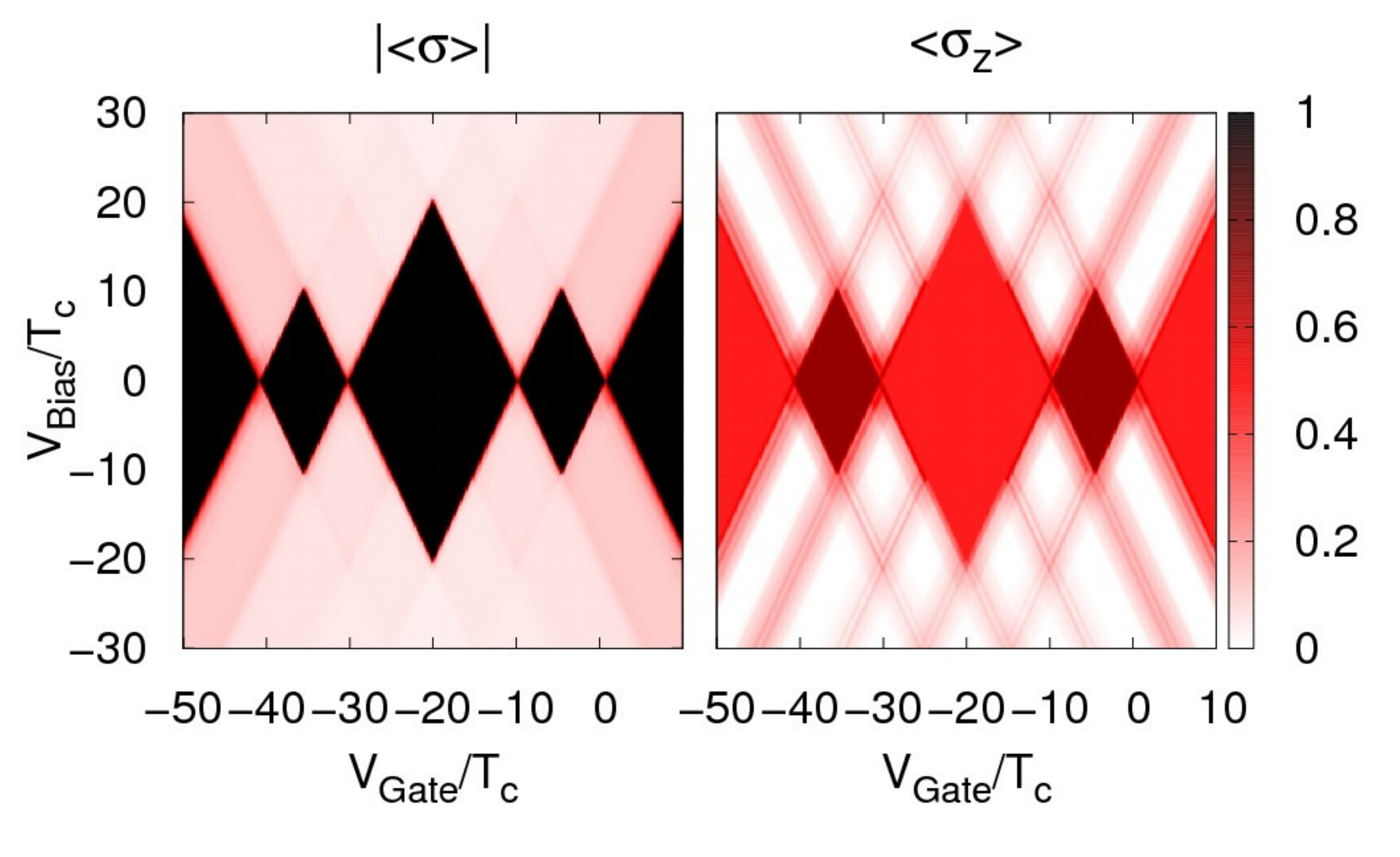}
	\caption{(Color online) State analysis of the charge qubit for low temperatures $\beta=100/T_{\rm{c}}$. The left picture shows the modulus of the Bloch vector and the right picture shows its $\sigma_z$ component. The system parameters are fixed to $U_\parallel=20\,T_{\rm{c}}$, $U_\perp=U_\vartimes=10\,T_{\rm{c}}$, $\Gamma_{i,\alpha}=J=\Omega=T_{\rm{c}}$ and $G=-T_{\rm{c}}$.}\label{F:CQBPurityAnalysis}
\end{figure}
The observation that there exist regions where the whole system is in a pure state leads to the question if the same is true for the CQB. A pure system density matrix together with a mixed CQB density matrix would imply entanglement between the transport DQDs and the CQB. In order to answer this question, we calculate the stationary expectation values of the CQB occupation operators according to $\left< \hat{\sigma}_{j} \right>=\rm{Tr}\left\{\hat{\sigma}_{\mathit{j}}\,\bar\rho \right\}$. Since the CQB represents a two-level system, these expectation values can be interpreted as the components of a Bloch vector. A pure CQB state corresponds to a Bloch vector with modulus one. Because in our model there is no $\hat{\sigma}_{\rm y}$ component for the CQB, the modulus of the Bloch vector is defined by $\|\left<\hat{\boldsymbol{\sigma}}\right>\|=\sqrt{\left<\hat{\sigma}_{\rm x}\right>^2+\left<\hat{\sigma}_{\rm z}\right>^2}$. We find that it is possible to tune the CQB to a pure qubit state for suitable combinations
of gate and bias voltage (see Fig.~\ref{F:CQBPurityAnalysis}). The regions with large CQB purity coincide with the Coulomb diamonds indicating that the CQB is not entangled with the rest of the system. Furthermore, there are well defined regions outside of the Coulomb diamonds where the CQB behaves completely classical, indicated by the vanishing of the Bloch vector modulus. However, as depicted in Fig.~\ref{F:CQBPurityAnalysis} the eigenstates that the CQB can take on within the Coulomb diamonds, do not correspond to localized electron states as the $\hat{\sigma}_z$-component never takes on the value $\pm 1$ in these regions. These properties remain qualitatively unchanged if other parameters are considered.
In general, we find that the purity properties are sensitive to thermal fluctuations and thus the stability of the corresponding regions will decrease as the temperature increases. This effect is strongest for the $2$-electron CD and weakest for the $4$-electron and $0$-electron CDs.
%
%
%%%%%%%%%%%%%%%%%%%%%%%%%%%%%%%%%%%%%%%%%%%%%%%%%%%%
\section{Entanglement}\label{S:Entanglement}    %%%%
%%%%%%%%%%%%%%%%%%%%%%%%%%%%%%%%%%%%%%%%%%%%%%%%%%%%
%
%
In contrast to the separable eigenstates in Eq.~\eqref{E:Eigenstates}, we can also change the Hamiltonian parameters to stabilize entangled states in the $2$-electron CD. Thus, motivated by investigations of the entanglement of the system \eqref{E:SysHamiltonian} without the CQB impurity \cite{Jordan2004,Clive2009}, we explore the effect of the presence of the CQB on the entanglement of the transport channels within the following section.\\
In order to qualitatively and quantitatively determine the entanglement between the two transport DQDs, we project the steady-state density matrix onto the two-qubit subspace where exactly one electron is present in each transport DQD. Subsequently, tracing out the CQB degrees of freedom yields an effective $4\times4$ matrix $\rho_2=\rm{Tr}_{\rm{rest}}\left\{\rho\right\}$ for the two coupled qubits represented by the transport DQDs. Here, the trace over ``rest" includes the $0$-, $1$-, $3$-, and $4$-electron subspaces as well as the $2$-electron states corresponding to a doubly occupied transport DQD. For this effective system of two coupled qubits, there exists a well-known entanglement measure \cite{Wotters1998}: The \textit{concurrence} $C=\max\left[0,\sqrt{\lambda_{1}} - \sum_{i=2}^{4} \sqrt{\lambda_{i}} \right]$ where $\lambda_{i}$ are the eigenvalues of $\rho_2 ( \sigma_y \otimes \sigma_y ) \rho_2^T ( \sigma_y \otimes \sigma_y )$ arranged in decreasing order, i.~e.~, $\lambda_{i+1}<\lambda_{i}$.
\subsection{Eigenstate Concurrence}
\begin{figure}[t]
  \subfigure{\centering\includegraphics[width=.478 \columnwidth]{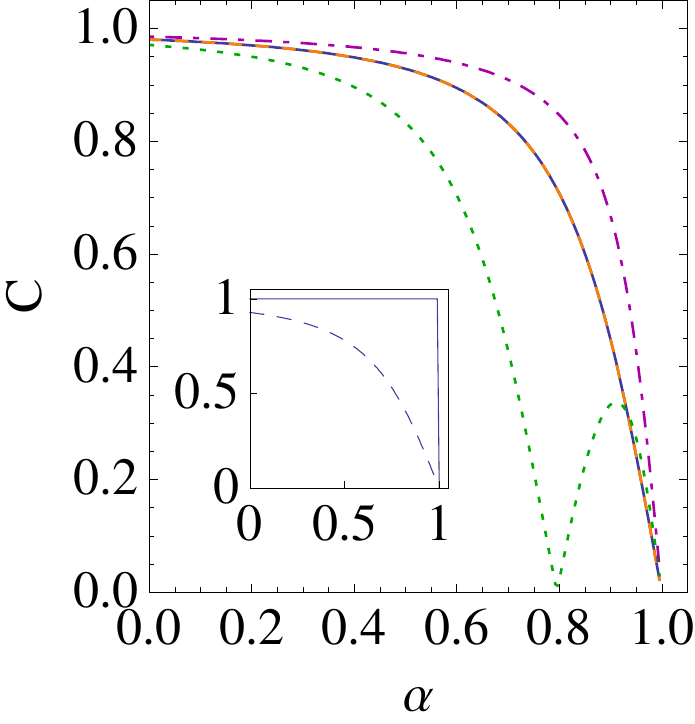}}
  \subfigure{\centering \includegraphics[width=0.509\columnwidth]{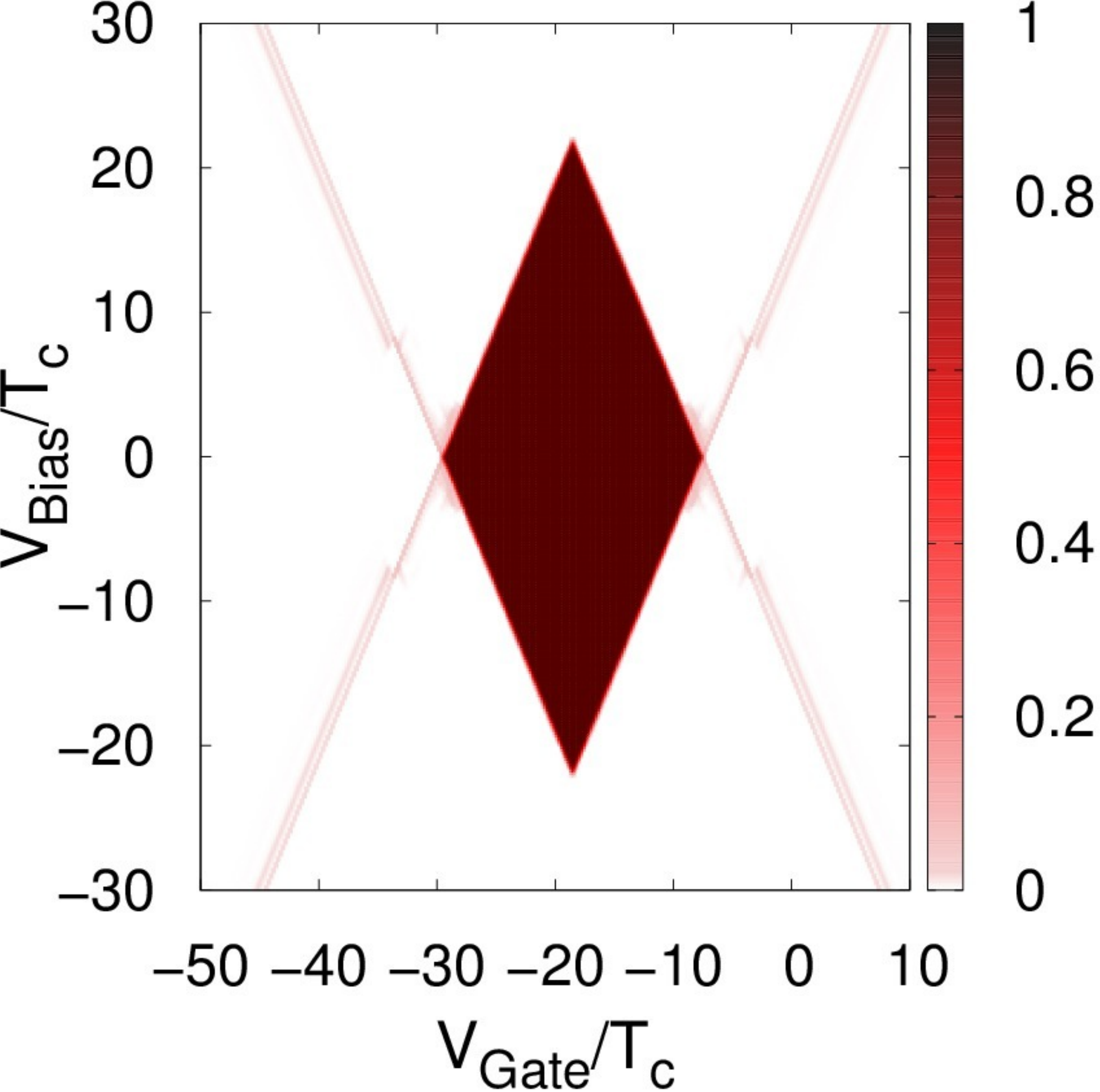}}
  \caption{(Color online) The left picture shows the concurrence of the four energy eigenstates in the qubit sector of Hamiltonian \eqref{E:SysHamiltonian} as a function of the ratio $\alpha$ of the Coulomb interaction strengths. For comparison the inset shows the concurrence for the system without CQB impurity. The right plot shows the generalized transport concurrence with $\alpha=0.7$ for the full transport spectrum. All parameters are fixed to $\beta=100/T_{\rm{c}}$, $U_\parallel=20\,T_{\rm{c}}$, $U_\perp=10\,T_{\rm{c}}$, $\Gamma_{i,\alpha}=J=\Omega=T_{\rm{c}}$ and $G=-T_{\rm{c}}$.}\label{F:EntanglementSpectra}
\end{figure}
For the system of two coupled transport DQDs it is known that the strength of entanglement depends on the strength of the on-site Coulomb interaction \cite{Clive2009}. Additionally, the entanglement of two parallel DQDs is also induced by asymmetric interactions. Hence, this property strongly depends on the choice of Coulomb interactions between the quantum dots. In our numerical evaluation in Fig.~\ref{F:FullTransportSpektrum} we explicitly assume that the Coulomb interaction between parallel dots has the same strength as the diagonal interactions, i.~e.~, $U_\vartimes=U_\perp$. However, with this choice, entanglement can hardly be achieved. In order to investigate the dependence of the entanglement of the pure states on the interaction strength, we introduce the ratio $\alpha=U_\vartimes / U_\perp$ between parallel and diagonal Coulomb interaction in the system Hamiltonian \eqref{E:SysHamiltonian}. First, investigating the concurrence of the pure qubit eigenstates for parallel DQDs with and without CQB in
dependence of this ratio $\alpha$ yields the results presented the left picture in Fig.~\ref{F:EntanglementSpectra}. In general we find that for the highly asymmetric case of $\alpha=0$ the concurrence for parallel DQDs with CQB impurity is maximal and reaches almost $1$. With increasing ratio $\alpha$ the concurrence decreases until it completely vanishes at $\alpha=1$, the point of maximum symmetry. In contrast, the two Bell states $ 1 /  \sqrt{2} \left( \ket{R,R} - \ket{L,L} \right) $ and $ 1 / \sqrt{2} \left( \ket{R,L} - \ket{L,R} \right) $ corresponding to the straight line in the inset in Fig.~\ref{F:EntanglementSpectra} have a constant concurrence of $1$ over almost the full range of $\alpha$. Hence, these eigenstates are maximally entangled states except near the point $\alpha=1$ where the concurrence vanishes discontinuously.  Note that always two of the four qubit states show exactly the same concurrence. Thus, both the solid and dashed lines in the inset correspond to two eigenstates each.
Comparing with the concurrence for qubit eigenstates of Eq.~\eqref{E:SysHamiltonian}, we find that the Bell states are destroyed due to the influence of the tunneling amplitude $J$ and the detuning $\Omega$ of the CQB. This result is in agreement with recent theoretical work using a configuration interaction method to analyze impurity effects on quantum bits \cite{Nguyen2011}. In addition, we observe that in general the concurrence of non-Bell states is raised for all values of $\alpha$. This effect mainly results from the modification parameter $G$. The concurrence is maximally enhanced for $G=-T_c$ and approaches the value of the unperturbed system as $G$ approaches zero.
\subsection{Generalized Transport Concurrence}
Next, we investigate the concurrence for the whole transport spectrum, which includes mixed states as well. Since both cases with $\alpha=0$ and $1$ are experimentally hard to achieve and also represent rather special configurations, we choose $\alpha=0.7$ for numerical investigations.\\
From the definition of the concurrence, it is obvious that this quantity is linear in the norm of the reduced density matrix $\rho_2$. Hence, if the reduced density matrix is not renormalized after tracing out the CQB, the concurrence is only exact if evaluated in the transport qubit sector where a single electron is on each transport DQD. Otherwise, the calculation yields a product of the exact concurrence multiplied by the probability for being in a transport qubit state. Due to this special property we will in the following use the non-renormalized concurrence $C$ to characterize the entanglement of the DQDs in the full gate and bias regimes. In the right picture of Fig.~\ref{F:EntanglementSpectra}, we plot the resulting concurrence versus gate and bias voltage.\\
We observe that the presence of the CQB enlarges the concurrence within the $2$-electron Coulomb diamond and slightly decreases the concurrence of the mixed states outside of this Coulomb diamond. Moreover, we see that in the exterior of the $2$-electron Coulomb diamond, the concurrence vanishes almost everywhere except for the regions associated with transitions from the $1$- or $3$-electron to the $2$-electron subspace. Here, the transport qubit eigenstates enter the steady-state reduced density matrix contributing their high concurrence to the mixture of states. However, since we do not re-normalize the effective two-qubit density matrix, this high concurrence gets multiplied by the $2$-electron fraction of the mixed state which corresponds to the probability to be in such a $2$-particle state. Going deeper into the $2$-electron subspace allows for more $2$-electron eigenstates to mix in the steady state reduced density matrix and hence rapidly reduces the respective entanglement. In an experiment the DQD,
entanglement could be measured for example via construction of a Bell inequality as suggested in Ref. \cite{Clive2009}.\\
%
%
%%%%%%%%%%%%%%%%%%%%%%%%%
\section{Summary}    %%%%
%%%%%%%%%%%%%%%%%%%%%%%%%
%
%
We studied the electronic transport properties of a system of two parallel DQDs which are subject to a CQB impurity that affects the transport through these DQDs. In particular, we investigated its effect on the steady-state currents in the strong and ultra-strong Coulomb blockade regimes. We find that the CQB detuning gives rise to an asymmetry that prevents the closing of Coulomb diamonds. Width as well as the position of the gap depend on the CQB parameters. In the ultra-strong Coulomb blockade regime we were able to extract analytic results for both, the steady-state current in the high-bias regime as well as the width and position of the gap in dependence of the CQB parameters. Moreover, we analyzed the purity of the reduced system density matrix and found that the purity of the system is preserved. In addition, the back-action on the CQB allows us to even render its eigenstates pure. Finally, we analyzed the impact of the CQB on the entanglement between the  two transport channels. Here, we observe on the
one hand a destruction of Bell states due to the CQB tunneling amplitude $J$ and detuning $\Omega$. On the other hand, we find an increase of entanglement of the remaining eigenstates due to the modification parameter $G$. In the exterior of the Coulomb diamond, entanglement is slightly decreased by the CQB.\\
Financial support by the DFG (SFB 910, SCHA 1646/2-1) is gratefully acknowledged. We would also like to thank S. Kohler for helpful discussions.
\appendix
%
%%%%%%%%%%%%%%%%%%%%%%%%%%%%%%%%%%%%%%%%%%%%%%%%%%%%%%%%
\section{Explicit Liouvillian}\label{S:Appendix}    %%%%
%%%%%%%%%%%%%%%%%%%%%%%%%%%%%%%%%%%%%%%%%%%%%%%%%%%%%%%%
%
The system Hamiltonian defined in Eq.~\eqref{E:SysHamiltonian} has a block structure in the local basis $\left\{\ket{\rm{DQD}_1,\rm{DQD}_2,\rm{CQB}}\right\}$ where each block is associated with a fixed number of electrons in the transport DQDs. In order to derive the explicit Liouvillian we first diagonalize the full system Hamiltonian \eqref{E:SysHamiltonian}
\begin{equation}
   \hat{H}_{\rm{S}}\ket{n,\alpha}=\varepsilon_{n,\alpha} \ket{n,\alpha}.
\end{equation}
Here $n \in \left\{ 0,1,2,3,4\right\}$ labels the number of electrons in the transport channels. The second index $\alpha \in \left\{ 1, \ldots, \alpha^{\rm{max}}_{n} \right\}$ with $\alpha^{\rm{max}}_{0}=\alpha^{\rm{max}}_{4}=2$, $\alpha^{\rm{max}}_{1}=\alpha^{\rm{max}}_{3}=8$ and $\alpha^{\rm{max}}_{2}=12$ labels the different eigenstates belonging to the same electronic subspace $n$. For convenience we first sort the basis vectors $\left\{\ket{n,\alpha}\right\}$ of the eigenbasis of $\hat{H}_{\rm{S}}$ by their increasing electron number $n$. Subsequently, the eigenstates belonging to the same number $n$ of electrons are sorted by their increasing eigenenergies $\varepsilon_{n,\alpha}$. This ordering leads to a Liouvillian with a compact block structure.\\
Introducing the abbreviation $\rho_{aa}= \bra{a} \rho \ket{a}$ for the elements of the density matrix in the energy eigenbasis $\ket{a}:=\ket{n,\alpha}$ allows to write down the effective rate equation for the populations \cite{Gernot2011} in the form
\begin{equation}% rate equation
   \dot\rho_{a a}=\sum_b \gamma_{ab,ab} \rho_{bb} - \left[ \sum_b \gamma_{ba,ba} \right] \rho_{aa}.
\end{equation}
The transition rates from state $b$ to state $a$ are defined via
\begin{equation}% transition rates
   \gamma_{ab,ab}=\sum_{\alpha, \beta=1}^8 \gamma_{\alpha \beta} \left( \varepsilon_b - \varepsilon_a \right) \bra{a}\hat{A}_{\alpha}\ket{b} \bra{a} \hat{A}^{\dagger}_{\beta}\ket{b}^*,\label{A:TransitionRates}
\end{equation}
with transition matrix elements involving the system operators
\begin{align}% system operators
&\hat{A}_1 = \hat{d}_{1,L},\, \hat{A}_2=\hat{d}_{1,L}^\dagger,\, \hat{A}_3 = \hat{d}_{1,R},\, \hat{A}_4=\hat{d}_{1,R}^\dagger,\notag\\
&\hat{A}_5 = \hat{d}_{2,L},\, \hat{A}_6=\hat{d}_{2,L}^\dagger,\, \hat{A}_7 = \hat{d}_{2,R},\, \hat{A}_8=\hat{d}_{2,R}^\dagger.
\end{align}
The coefficients $\gamma_{\alpha \beta}(\omega)$ in Eq.~\eqref{A:TransitionRates} correspond to the even Fourier transformed bath correlation functions
\begin{equation}% bath correlation functions
   \gamma_{\alpha \beta}(\omega)=\int_{-\infty}^{+\infty} \rm{Tr}\left\{e^{\ii \mathit{\hat{H}}_{\rm{B}} \tau} \mathit{\hat{B}}_\alpha e^{-\ii \mathit{\hat{H}}_{\rm{B}} \tau} \mathit{\hat{B}}_\beta \, \bar\rho_{\rm{B}}\right\} e^{\rm{i} \omega \tau}d\tau,
\end{equation}
which involve the bath operators
\begin{align}% bath operators
&\hat{B}_1 = \sum_k t_{1,L,k}\, \hat{q}_{1,L,k}^\dagger,\, \hat{B}_2 = \sum_k t_{1,L,k}^*\, \hat{q}_{1,L,k},\notag\\
&\hat{B}_3 = \sum_k t_{1,R,k}\, \hat{q}_{1,R,k}^\dagger,\, \hat{B}_4 = \sum_k t_{1,R,k}^*\, \hat{q}_{1,R,k},\notag\\
&\hat{B}_5 = \sum_k t_{2,L,k}\, \hat{q}_{2,L,k}^\dagger,\, \hat{B}_6 = \sum_k t_{2,L,k}^*\, \hat{q}_{2,L,k},\notag\\
&\hat{B}_7 = \sum_k t_{2,R,k}\, \hat{q}_{2,R,k}^\dagger,\, \hat{B}_8 = \sum_k t_{2,R,k}^*\, \hat{q}_{2,R,k}.
\end{align}
From this definition follows that the non-vanishing coefficients are given by
\begin{align}
   \gamma_{12}\left(\omega\right)&=\Gamma_{1,L}\left(-\omega\right)f_{1,L}\left(-\omega\right),\notag\\
   \gamma_{21}\left(\omega\right)&=\Gamma_{1,L}\left(\omega\right)\left[1-f_{1,L}\left(\omega\right)\right],\notag\\
   \gamma_{34}\left(\omega\right)&=\Gamma_{1,R}\left(-\omega\right)f_{1,R}\left(-\omega\right),\notag\\
   \gamma_{43}\left(\omega\right)&=\Gamma_{1,R}\left(\omega\right)\left[1-f_{1,R}\left(\omega\right)\right],\notag\\
   \gamma_{56}\left(\omega\right)&=\Gamma_{2,L}\left(-\omega\right)f_{2,L}\left(-\omega\right),\notag\\
   \gamma_{65}\left(\omega\right)&=\Gamma_{2,L}\left(\omega\right)\left[1-f_{2,L}\left(\omega\right)\right],\notag\\
   \gamma_{78}\left(\omega\right)&=\Gamma_{2,R}\left(-\omega\right)f_{2,R}\left(-\omega\right),\notag\\
   \gamma_{87}\left(\omega\right)&=\Gamma_{2,R}\left(\omega\right)\left[1-f_{2,R}\left(\omega\right)\right].
\end{align}
%
%
%%%%%%%%%%%%%%%%%%%%%%%
%%%% BIBLIOGRAPHIE %%%%
%%%%%%%%%%%%%%%%%%%%%%%
%
%
%\bibliography{references}

\begin{thebibliography}{10}%
\makeatletter
\providecommand \@ifxundefined [1]{%
 \ifx #1\undefined \expandafter \@firstoftwo
 \else \expandafter \@secondoftwo
\fi
}%
\providecommand \@ifnum [1]{%
 \ifnum #1\expandafter \@firstoftwo
 \else \expandafter \@secondoftwo
\fi
}%
\providecommand \enquote [1]{``#1''}%
\providecommand \bibnamefont  [1]{#1}%
\providecommand \bibfnamefont [1]{#1}%
\providecommand \citenamefont [1]{#1}%
\providecommand\href[0]{\@sanitize\@href}%
\providecommand\@href[1]{\endgroup\@@startlink{#1}\endgroup\@@href}%
\providecommand\@@href[1]{#1\@@endlink}%
\providecommand \@sanitize [0]{\begingroup\catcode`\&12\catcode`\#12\relax}%
\@ifxundefined \pdfoutput {\@firstoftwo}{%
 \@ifnum{\z@=\pdfoutput}{\@firstoftwo}{\@secondoftwo}%
}{%
 \providecommand\@@startlink[1]{\leavevmode}%
 \providecommand\@@endlink[0]{}%
}{%
 \providecommand\@@startlink[1]{%
  \leavevmode
  \pdfstartlink
   attr{/Border[0 0 1 ]/H/I/C[0 1 1]}%
   user{/Subtype/Link/A<</Type/Action/S/URI/URI(#1)>>}%
  \relax
 }%
 \providecommand\@@endlink[0]{\pdfendlink}%
}%
\providecommand \url  [0]{\begingroup\@sanitize \@url }%
\providecommand \@url [1]{\endgroup\@href {#1}{\urlprefix}}%
\providecommand \urlprefix [0]{URL }%
\providecommand \Eprint[0]{\href }%
\@ifxundefined \urlstyle {%
  \providecommand \doi [1]{doi:\discretionary{}{}{}#1}%
}{%
  \providecommand \doi [0]{doi:\discretionary{}{}{}\begingroup
  \urlstyle{rm}\Url }%
}%
\providecommand \doibase [0]{http://dx.doi.org/}%
\providecommand \Doi[1]{\href{\doibase#1}}%
\providecommand \bibAnnote [3]{%
  \BibitemShut{#1}%
  \begin{quotation}\noindent
    \textsc{Key:}\ #2\\\textsc{Annotation:}\ #3%
  \end{quotation}%
}%
\providecommand \bibAnnoteFile [2]{%
  \IfFileExists{#2}{\bibAnnote {#1} {#2} {\input{#2}}}{}%
}%
\providecommand \typeout [0]{\immediate \write \m@ne }%
\providecommand \selectlanguage [0]{\@gobble}%
\providecommand \bibinfo [0]{\@secondoftwo}%
\providecommand \bibfield [0]{\@secondoftwo}%
\providecommand \translation [1]{[#1]}%
\providecommand \BibitemOpen[0]{}%
\providecommand \bibitemStop [0]{}%
\providecommand \bibitemNoStop [0]{.\EOS\space}%
\providecommand \EOS [0]{\spacefactor3000\relax}%
\providecommand \BibitemShut [1]{\csname bibitem#1\endcsname}%
%</preamble>
%1
\bibitem{DiVincenco2000}%
  \BibitemOpen
  \bibfield{author}{%
  \bibinfo {author} {\bibfnamefont{D.~P.}\ \bibnamefont{DiVincenzo}},\ }%
  \bibfield{journal}{%
  {\bibinfo
  {journal} {Fortschr. Phys.}}\ }%
  \textbf{\bibinfo {volume} {48}},\ \bibinfo {pages} {771} (\bibinfo {year}
  {2000})%
  \bibAnnoteFile{NoStop}{DiVincenco2000}
%2
\bibitem{Loss1998}%
  \BibitemOpen
  \bibfield{author}{%
  \bibinfo {author} {\bibfnamefont{D.}~\bibnamefont{Loss}}\ and\ \bibinfo
  {author} {\bibfnamefont{D.~P.}\ \bibnamefont{DiVincenzo}},\ }%
  \bibfield{journal}{%
  {\bibinfo {journal} {Phys. Rev. A}}\ }%
  \textbf{\bibinfo {volume} {57}},\ \bibinfo {pages} {120} (\bibinfo {year} {1998})%
  \bibAnnoteFile{NoStop}{Loss1998}
%3
\bibitem{Blick1998}%
  \BibitemOpen
  \bibfield{author}{%
  \bibinfo {author} {\bibfnamefont{R.~H.}\ \bibnamefont{Blick}}, \bibinfo
  {author} {\bibfnamefont{D.}~\bibnamefont{Pfannkuche}}, \bibinfo {author}
  {\bibfnamefont{R.~J.}\ \bibnamefont{Haug}}, \bibinfo {author}
  {\bibfnamefont{K.~v.}\ \bibnamefont{Klitzing}},\ and\ \bibinfo {author}
  {\bibfnamefont{K.}~\bibnamefont{Eberl}},\ }%
  \bibfield{journal}{%
  {\bibinfo {journal} {Phys. Rev. Lett.}}\ }%
  \textbf{\bibinfo {volume} {80}},\ \bibinfo {pages} {4032} (\bibinfo {year} {1998})%
  \bibAnnoteFile{NoStop}{Blick1998}
%4
\bibitem{Vion2002}%
  \BibitemOpen
  \bibfield{author}{%
  \bibinfo {author} {\bibfnamefont{D.}~\bibnamefont{Vion}}, \bibinfo {author}
  {\bibfnamefont{A.}~\bibnamefont{Aassime}}, \bibinfo {author}
  {\bibfnamefont{A.}~\bibnamefont{Cottet}}, \bibinfo {author}
  {\bibfnamefont{P.}~\bibnamefont{Joyez}}, \bibinfo {author}
  {\bibfnamefont{H.}~\bibnamefont{Pothier}}, \bibinfo {author}
  {\bibfnamefont{C.}~\bibnamefont{Urbina}}, \bibinfo {author}
  {\bibfnamefont{D.}~\bibnamefont{Esteve}},\ and\ \bibinfo {author}
  {\bibfnamefont{M.~H.}\ \bibnamefont{Devoret}},\ }%
  \bibfield{journal}{%
  {\bibinfo {journal} {Science}}\ }%
  \textbf{\bibinfo {volume} {296}},\ \bibinfo {pages} {886} (\bibinfo {year}
  {2002})%
  \bibAnnoteFile{NoStop}{Vion2002}
%5
\bibitem{DiVincenco2005}%
  \BibitemOpen
  \bibfield{author}{%
  \bibinfo {author} {\bibfnamefont{D.~P.}\ \bibnamefont{DiVincenzo}},\ }%
  \bibfield{journal}{%
  {\bibinfo {journal} {Science}}\ }%
  \textbf{\bibinfo {volume} {309}},\ \bibinfo {pages} {2173} (\bibinfo {year}
  {2005})%
  \bibAnnoteFile{NoStop}{DiVincenco2005}
%6
\bibitem{Wiel2006}%
  \BibitemOpen
  \bibfield{author}{%
  \bibinfo {author} {\bibfnamefont{W.~G.}\ \bibnamefont{van~der Wiel}},
  \bibinfo {author} {\bibfnamefont{S.}~\bibnamefont{De~Franceschi}}, \bibinfo
  {author} {\bibfnamefont{J.~M.}\ \bibnamefont{Elzerman}}, \bibinfo {author}
  {\bibfnamefont{T.}~\bibnamefont{Fujisawa}}, \bibinfo {author}
  {\bibfnamefont{S.}~\bibnamefont{Tarucha}},\ and\ \bibinfo {author}
  {\bibfnamefont{L.~P.}\ \bibnamefont{Kouwenhoven}},\ }%
  \bibfield{journal}{%
  {\bibinfo {journal} {Rev. Mod. Phys.}}\ }%
  \textbf{\bibinfo {volume} {75}},\ \bibinfo {pages} {1} (\bibinfo {year} {2002})%
  \bibAnnoteFile{NoStop}{Wiel2006}
%7
\bibitem{Oosterkamp1998}%
  \BibitemOpen
  \bibfield{author}{%
  \bibinfo {author} {\bibfnamefont{T.~H.}\ \bibnamefont{Oosterkamp}}, \bibinfo
  {author} {\bibfnamefont{T.}~\bibnamefont{Fujisawa}}, \bibinfo {author}
  {\bibfnamefont{W.~G.}\ \bibnamefont{van~der Wiel}}, \bibinfo {author}
  {\bibfnamefont{K.}~\bibnamefont{Ishibashi}}, \bibinfo {author}
  {\bibfnamefont{R.~V.}\ \bibnamefont{Hijman}}, \bibinfo {author}
  {\bibfnamefont{S.}~\bibnamefont{Tarucha}},\ and\ \bibinfo {author}
  {\bibfnamefont{L.~P.}\ \bibnamefont{Kouwenhoven}},\ }%
  \bibfield{journal}{%
  \bibinfo {journal} {Nature}\ }%
  \textbf{\bibinfo {volume} {395}},\ \bibinfo {pages} {873} (\bibinfo {year}
  {1998})%
  \bibAnnoteFile{NoStop}{Oosterkamp1998}
%8
\bibitem{Petta2005}%
  \BibitemOpen
  \bibfield{author}{%
  \bibinfo {author} {\bibfnamefont{J.~R.}\ \bibnamefont{Petta}}, \bibinfo
  {author} {\bibfnamefont{A.~C.}\ \bibnamefont{Johnson}}, \bibinfo {author}
  {\bibfnamefont{J.~M.}\ \bibnamefont{Taylor}}, \bibinfo {author}
  {\bibfnamefont{E.~A.}\ \bibnamefont{Laird}}, \bibinfo {author}
  {\bibfnamefont{A.}~\bibnamefont{Yacoby}}, \bibinfo {author}
  {\bibfnamefont{M.~D.}\ \bibnamefont{Lukin}}, \bibinfo {author}
  {\bibfnamefont{C.~M.}\ \bibnamefont{Marcus}}, \bibinfo {author}
  {\bibfnamefont{M.~P.}\ \bibnamefont{Hanson}},\ and\ \bibinfo {author}
  {\bibfnamefont{A.~C.}\ \bibnamefont{Gossard}},\ }%
  \bibfield{journal}{%
  {\bibinfo {journal} {Science}}\ }%
  \textbf{\bibinfo {volume} {309}},\ \bibinfo {pages} {2180} (\bibinfo {year}
  {2005})%
  \bibAnnoteFile{NoStop}{Petta2005}
%9
\bibitem{Koppens2006}%
  \BibitemOpen
  \bibfield{author}{%
  \bibinfo {author} {\bibfnamefont{F.~H.~L.}\ \bibnamefont{Koppens}}, \bibinfo
  {author} {\bibfnamefont{C.}~\bibnamefont{Buizert}}, \bibinfo {author}
  {\bibfnamefont{K.~J.}\ \bibnamefont{Tielrooij}}, \bibinfo {author}
  {\bibfnamefont{I.~T.}\ \bibnamefont{Vink}}, \bibinfo {author}
  {\bibfnamefont{K.~C.}\ \bibnamefont{Nowack}}, \bibinfo {author}
  {\bibfnamefont{T.}~\bibnamefont{Meunier}}, \bibinfo {author}
  {\bibfnamefont{L.~P.}\ \bibnamefont{Kouwenhoven}},\ and\ \bibinfo {author}
  {\bibfnamefont{L.~M.~K.}\ \bibnamefont{Vandersypen}},\ }%
  \bibfield{journal}{%
  \bibinfo {journal} {Nature}\ }%
  \textbf{\bibinfo {volume} {442}},\ \bibinfo {pages} {766} (\bibinfo {year}
  {2006})%
  \bibAnnoteFile{NoStop}{Koppens2006}
%10
\bibitem{Stehlik2011}%
  \BibitemOpen
  \bibfield{author}{%
  \bibinfo {author} {\bibfnamefont{Y.}~\bibnamefont{Dovzhenko}}, \bibinfo
  {author} {\bibfnamefont{J.}~\bibnamefont{Stehlik}}, \bibinfo {author}
  {\bibfnamefont{K.~D.}\ \bibnamefont{Petersson}}, \bibinfo {author}
  {\bibfnamefont{J.~R.}\ \bibnamefont{Petta}}, \bibinfo {author}
  {\bibfnamefont{H.}~\bibnamefont{Lu}},\ and\ \bibinfo {author}
  {\bibfnamefont{A.~C.}\ \bibnamefont{Gossard}},\ }%
  \bibfield{journal}{%
  {\bibinfo {journal} {Phys. Rev. B}}\ }%
  \textbf{\bibinfo {volume} {84}},\ \bibinfo {pages} {161302} (\bibinfo {year} {2011})%
  \bibAnnoteFile{NoStop}{Stehlik2011}
%11
\bibitem{Barthel2009}%
  \BibitemOpen
  \bibfield{author}{%
  \bibinfo {author} {\bibfnamefont{C.}~\bibnamefont{Barthel}}, \bibinfo
  {author} {\bibfnamefont{D.~J.}\ \bibnamefont{Reilly}}, \bibinfo {author}
  {\bibfnamefont{C.~M.}\ \bibnamefont{Marcus}}, \bibinfo {author}
  {\bibfnamefont{M.~P.}\ \bibnamefont{Hanson}},\ and\ \bibinfo {author}
  {\bibfnamefont{A.~C.}\ \bibnamefont{Gossard}},\ }%
  \bibfield{journal}{%
  {\bibinfo {journal} {Phys. Rev. Lett.}}\
  }%
  \textbf{\bibinfo {volume} {103}},\ \bibinfo {pages} {160503} (\bibinfo {year} {2009})%
  \bibAnnoteFile{NoStop}{Barthel2009}
%12
\bibitem{Petersson2010}%
  \BibitemOpen
  \bibfield{author}{%
  \bibinfo {author} {\bibfnamefont{K.~D.}\ \bibnamefont{Petersson}}, \bibinfo
  {author} {\bibfnamefont{J.~R.}\ \bibnamefont{Petta}}, \bibinfo {author}
  {\bibfnamefont{H.}~\bibnamefont{Lu}},\ and\ \bibinfo {author}
  {\bibfnamefont{A.~C.}\ \bibnamefont{Gossard}},\ }%
  \bibfield{journal}{%
  {\bibinfo {journal} {Phys. Rev. Lett.}}\
  }%
  \textbf{\bibinfo {volume} {105}},\ \bibinfo {pages} {246804} (\bibinfo {year} {2010})%
  \bibAnnoteFile{NoStop}{Petersson2010}
%13
\bibitem{Nowack2011}%
  \BibitemOpen
  \bibfield{author}{%
  \bibinfo {author} {\bibfnamefont{K.~C.}\ \bibnamefont{Nowack}}, \bibinfo
  {author} {\bibfnamefont{M.}~\bibnamefont{Shafiei}}, \bibinfo {author}
  {\bibfnamefont{M.}~\bibnamefont{Laforest}}, \bibinfo {author}
  {\bibfnamefont{G.~E. D.~K.}\ \bibnamefont{Prawiroatmodjo}}, \bibinfo {author}
  {\bibfnamefont{L.~R.}\ \bibnamefont{Schreiber}}, \bibinfo {author}
  {\bibfnamefont{C.}~\bibnamefont{Reichl}}, \bibinfo {author}
  {\bibfnamefont{W.}~\bibnamefont{Wegscheider}},\ and\ \bibinfo {author}
  {\bibfnamefont{L.~M.~K.}\ \bibnamefont{Vandersypen}},\ }%
  \bibfield{journal}{%
  {\bibinfo {journal} {Science}}\ }%
  \textbf{\bibinfo {volume} {333}},\ \bibinfo {pages} {1269} (\bibinfo {year}
  {2011})%
  \bibAnnoteFile{NoStop}{Nowack2011}
%14
\bibitem{Kohler2005}%
  \BibitemOpen
  \bibfield{author}{%
  \bibinfo {author} {\bibfnamefont{M.~J.}\ \bibnamefont{Storcz}}, \bibinfo
  {author} {\bibfnamefont{U.}~\bibnamefont{Hartmann}}, \bibinfo {author}
  {\bibfnamefont{S.}~\bibnamefont{Kohler}},\ and\ \bibinfo {author}
  {\bibfnamefont{F.~K.}\ \bibnamefont{Wilhelm}},\ }%
  \bibfield{journal}{%
  {\bibinfo {journal} {Phys. Rev. B}}\ }%
  \textbf{\bibinfo {volume} {72}},\ \bibinfo {pages} {235321} (\bibinfo {year} {2005})%
  \bibAnnoteFile{NoStop}{Kohler2005}
%15
\bibitem{Kohler2005a}%
  \BibitemOpen
  \bibfield{author}{%
  \bibinfo {author} {\bibfnamefont{K.~M.}\ \bibnamefont{Fonseca-Romero}},
  \bibinfo {author} {\bibfnamefont{S.}~\bibnamefont{Kohler}},\ and\ \bibinfo
  {author} {\bibfnamefont{P.}~\bibnamefont{H\"anggi}},\ }%
  \bibfield{journal}{%
  {\bibinfo {journal} {Phys. Rev. Lett.}}\
  }%
  \textbf{\bibinfo {volume} {95}},\ \bibinfo {pages} {140502} (\bibinfo {year} {2005})%
  \bibAnnoteFile{NoStop}{Kohler2005a}
%16
\bibitem{Kohler2009}%
  \BibitemOpen
  \bibfield{author}{%
  \bibinfo {author} {\bibfnamefont{D.}~\bibnamefont{Zueco}}, \bibinfo {author}
  {\bibfnamefont{F.}~\bibnamefont{Galve}}, \bibinfo {author}
  {\bibfnamefont{S.}~\bibnamefont{Kohler}},\ and\ \bibinfo {author}
  {\bibfnamefont{P.}~\bibnamefont{H\"anggi}},\ }%
  \bibfield{journal}{%
  {\bibinfo {journal} {Phys. Rev. A}}\ }%
  \textbf{\bibinfo {volume} {80}},\ \bibinfo {pages} {042303} (\bibinfo {year} {2009})%
  \bibAnnoteFile{NoStop}{Kohler2009}
%17
\bibitem{Trauzettel2006}%
  \BibitemOpen
  \bibfield{author}{%
  \bibinfo {author} {\bibfnamefont{B.}~\bibnamefont{Trauzettel}}, \bibinfo
  {author} {\bibfnamefont{A.~N.}\ \bibnamefont{Jordan}}, \bibinfo {author}
  {\bibfnamefont{C.~W.~J.}\ \bibnamefont{Beenakker}},\ and\ \bibinfo {author}
  {\bibfnamefont{M.}~\bibnamefont{B\"uttiker}},\ }%
  \bibfield{journal}{%
  {\bibinfo {journal} {Phys. Rev. B}}\ }%
  \textbf{\bibinfo {volume} {73}},\ \bibinfo {pages} {235331} (\bibinfo {year} {2006})%
  \bibAnnoteFile{NoStop}{Trauzettel2006}
%18
\bibitem{Clive2009}%
  \BibitemOpen
  \bibfield{author}{%
  \bibinfo {author} {\bibfnamefont{C.}~\bibnamefont{Emary}},\ }%
  \bibfield{journal}{%
  {\bibinfo {journal} {Phys. Rev. B}}\ }%
  \textbf{\bibinfo {volume} {80}},\ \bibinfo {pages} {161309} (\bibinfo {year} {2009})%
  \bibAnnoteFile{NoStop}{Clive2009}
%19
\bibitem{Brandes2007}%
  \BibitemOpen
  \bibfield{author}{%
  \bibinfo {author} {\bibfnamefont{C.-M.}\ \bibnamefont{Li}}, \bibinfo {author}
  {\bibfnamefont{L.-Y.}\ \bibnamefont{Hsu}}, \bibinfo {author}
  {\bibfnamefont{Y.-N.}\ \bibnamefont{Chen}}, \bibinfo {author}
  {\bibfnamefont{D.-S.}\ \bibnamefont{Chuu}},\ and\ \bibinfo {author}
  {\bibfnamefont{T.}~\bibnamefont{Brandes}},\ }%
  \bibfield{journal}{%
  {\bibinfo {journal} {Phys. Rev. A}}\ }%
  \textbf{\bibinfo {volume} {76}},\ \bibinfo {pages} {032313} (\bibinfo {year} {2007})%
  \bibAnnoteFile{NoStop}{Brandes2007}
%20
\bibitem{Reuther2011}%
  \BibitemOpen
  \bibfield{author}{%
  \bibinfo {author} {\bibfnamefont{G.~M.}\ \bibnamefont{Reuther}}, \bibinfo
  {author} {\bibfnamefont{D.}~\bibnamefont{Zueco}}, \bibinfo {author}
  {\bibfnamefont{P.}~\bibnamefont{H\"anggi}},\ and\ \bibinfo {author}
  {\bibfnamefont{S.}~\bibnamefont{Kohler}},\ }%
  \bibfield{journal}{%
  {\bibinfo {journal} {Phys. Rev. B}}\ }%
  \textbf{\bibinfo {volume} {83}},\ \bibinfo {pages} {014303} (\bibinfo {year} {2011})%
  \bibAnnoteFile{NoStop}{Reuther2011}
%21
\bibitem{Kouwenhoven1996}%
  \BibitemOpen
  \bibfield{author}{%
  \bibinfo {author} {\bibfnamefont{S.}~\bibnamefont{Tarucha}}, \bibinfo
  {author} {\bibfnamefont{D.~G.}\ \bibnamefont{Austing}}, \bibinfo {author}
  {\bibfnamefont{T.}~\bibnamefont{Honda}}, \bibinfo {author}
  {\bibfnamefont{R.~J.}\ \bibnamefont{van~der Hage}},\ and\ \bibinfo {author}
  {\bibfnamefont{L.~P.}\ \bibnamefont{Kouwenhoven}},\ }%
  \bibfield{journal}{%
  {\bibinfo {journal} {Phys. Rev. Lett.}}\ }%
  \textbf{\bibinfo {volume} {77}},\ \bibinfo {pages} {3613} (\bibinfo {year} {1996})%
  \bibAnnoteFile{NoStop}{Kouwenhoven1996}
%22
\bibitem{Kouwenhoven1997}%
  \BibitemOpen
  \bibfield{author}{%
  \bibinfo {author} {\bibfnamefont{L.~P.}\ \bibnamefont{Kouwenhoven}}, \bibinfo
  {author} {\bibfnamefont{C.~M.}\ \bibnamefont{Marcus}}, \bibinfo {author}
  {\bibfnamefont{P.~L.}\ \bibnamefont{McEuen}}, \bibinfo {author}
  {\bibfnamefont{S.}~\bibnamefont{Tarucha}}, \bibinfo {author}
  {\bibfnamefont{R.~M.}\ \bibnamefont{Westervelt}},\ and\ \bibinfo {author}
  {\bibfnamefont{N.~S.}\ \bibnamefont{Wingreen}},\ }%
  in\ \emph{\bibinfo {booktitle} {Mesoscopic Electron Transport}},\ \bibinfo
  {series} {Series E: Applied Sciences}, Vol.\ \bibinfo {volume} {345},\
  \bibinfo {editor} {edited by\ \bibinfo {editor} {\bibfnamefont{L.~L.}\
  \bibnamefont{Sohn}}, \bibinfo {editor} {\bibfnamefont{L.~P.}\
  \bibnamefont{Kouwenhoven}},\ and\ \bibinfo {editor}
  {\bibfnamefont{G.}~\bibnamefont{Schon}}}\ (\bibinfo {publisher} {Kluwer
  Academic Publishers, Dordrecht/Boston/London},\ \bibinfo {year} {1997})\ pp.\
  \bibinfo {pages} {105--214}%
  \bibAnnoteFile{NoStop}{Kouwenhoven1997}
%23
\bibitem{Klein1996}%
  \BibitemOpen
  \bibfield{author}{%
  \bibinfo {author} {\bibfnamefont{D.~L.}\ \bibnamefont{Klein}}, \bibinfo
  {author} {\bibfnamefont{P.~L.}\ \bibnamefont{McEuen}}, \bibinfo {author}
  {\bibfnamefont{J.~E.~B.}\ \bibnamefont{Katari}}, \bibinfo {author}
  {\bibfnamefont{R.}~\bibnamefont{Roth}},\ and\ \bibinfo {author}
  {\bibfnamefont{A.~P.}\ \bibnamefont{Alivisatos}},\ }%
  \bibfield{journal}{%
  {\bibinfo {journal} {Appl. Phys. Lett.}}\ }%
  \textbf{\bibinfo {volume} {68}},\ \bibinfo {pages} {2574} (\bibinfo {year}
  {1996})%
  \bibAnnoteFile{NoStop}{Klein1996}
%24
\bibitem{Elzerman2003}%
  \BibitemOpen
  \bibfield{author}{%
  \bibinfo {author} {\bibfnamefont{J.~M.}\ \bibnamefont{Elzerman}},
  \bibinfo {author} {\bibfnamefont{R.}~\bibnamefont{Hanson}},
  \bibinfo {author} {\bibfnamefont{J.~S.}\ \bibnamefont{Greidanus}},
  \bibinfo {author} {\bibfnamefont{L.~H.~Willems}\ \bibnamefont{van~Beveren}},
  \bibinfo {author} {\bibfnamefont{S.}~\bibnamefont{De~Franceschi}},
  \bibinfo {author} {\bibfnamefont{L.~M.~K.}\ \bibnamefont{Vandersypen}},
  \bibinfo {author} {\bibfnamefont{S.}~\bibnamefont{Tarucha}},\ and\
  \bibinfo {author} {\bibfnamefont{L.~P.}\ \bibnamefont{Kouwenhoven}},\ }%
  \bibfield{journal}{%
  {\bibinfo {journal} {Phys. Rev. B}}\ }%
  \textbf{\bibinfo {volume} {67}},\ \bibinfo {pages} {161308} (\bibinfo {year} {2003})%
  \bibAnnoteFile{NoStop}{Elzerman2003}
%25
\bibitem{Nguyen2011}%
  \BibitemOpen
  \bibfield{author}{%
  \bibinfo {author} {\bibfnamefont{N.~T.~T.}~\bibnamefont{Nguyen}}\ and\ \bibinfo
  {author} {\bibfnamefont{S.}~\bibnamefont{Das~Sarma}},\ }%
  \bibfield{journal}{%
  {\bibinfo {journal} {Phys. Rev. B}}\ }%
  \textbf{\bibinfo {volume} {83}},\ \bibinfo {pages} {235322} (\bibinfo {year} {2011})%
  \bibAnnoteFile{NoStop}{Nguyen2011}
%26
\bibitem{Barnes2011}%
  \BibitemOpen
  \bibfield{author}{%
  \bibinfo {author} {\bibfnamefont{E.}~\bibnamefont{Barnes}},
  \bibinfo {author} {\bibfnamefont{J.~P.}~\bibnamefont{Kestner}},
  \bibinfo {author} {\bibfnamefont{N.~T.~T.}\ \bibnamefont{Nguyen}},\ and\
  \bibinfo {author} {\bibfnamefont{S.}~\bibnamefont{Das~Sarma}},\ }%
  \bibfield{journal}{%
  \bibinfo {journal} {Phys. Rev. B}\ }%
  \textbf{\bibinfo {volume} {84}},\ \bibinfo {pages} {235309} (\bibinfo {year}
  {2011})%
  \bibAnnoteFile{NoStop}{Barnes2011}
%27
\bibitem{Chorley2011}%
  \BibitemOpen
  \bibfield{author}{%
  \bibinfo {author} {\bibfnamefont{S.~J.}~\bibnamefont{Chorley}},
  \bibinfo {author} {\bibfnamefont{G.}~\bibnamefont{Giavaras}},
  \bibinfo {author} {\bibfnamefont{J.}~\bibnamefont{Wabnig}},
  \bibinfo {author} {\bibfnamefont{G.~A.~C.}~\bibnamefont{Jones}},
  \bibinfo {author} {\bibfnamefont{C.~G.}~\bibnamefont{Smith}},
  \bibinfo {author} {\bibfnamefont{G.~A.~D.}~\bibnamefont{Briggs}},\ and\
  \bibinfo {author} {\bibfnamefont{M.~R.}~\bibnamefont{Buitelaar}},\ }%
  \bibfield{journal}{%
  {\bibinfo {journal} {Phys. Rev. Lett.}}\
  }%
  \textbf{\bibinfo {volume} {106}},\ \bibinfo {pages} {206801} (\bibinfo {year} {2011})%
  \bibAnnoteFile{NoStop}{Chorley2011}
%28
\bibitem{Kreisbeck2010}%
  \BibitemOpen
  \bibfield{author}{%
  \bibinfo {author} {\bibfnamefont{C.}~\bibnamefont{Kreisbeck}}, \bibinfo
  {author} {\bibfnamefont{F.~J.}\ \bibnamefont{Kaiser}},\ and\ \bibinfo
  {author} {\bibfnamefont{S.}~\bibnamefont{Kohler}},\ }%
  \bibfield{journal}{%
  {\bibinfo {journal} {Phys. Rev. B}}\ }%
  \textbf{\bibinfo {volume} {81}},\ \bibinfo {pages} {125404} (\bibinfo {year} {2010})%
  \bibAnnoteFile{NoStop}{Kreisbeck2010}
%29
\bibitem{Breuer}%
  \BibitemOpen
  \bibfield{author}{%
  \bibinfo {author} {\bibfnamefont{H.~P.}\ \bibnamefont{Breuer}}\ and\ \bibinfo
  {author} {\bibfnamefont{F.}~\bibnamefont{Petruccione}},\ }%
  \emph{\bibinfo {title} {The Theory of Open Quantum Systems}}\ (\bibinfo
  {publisher} {Oxford University Press, Oxford},\ \bibinfo {year} {2002})%
  \bibAnnoteFile{NoStop}{Breuer}
%30
\bibitem{Gernot2009}%
  \BibitemOpen
  \bibfield{author}{%
  \bibinfo {author} {\bibfnamefont{G.}~\bibnamefont{Schaller}}, \bibinfo
  {author} {\bibfnamefont{G.}~\bibnamefont{Kie\ss{}lich}},\ and\ \bibinfo
  {author} {\bibfnamefont{T.}~\bibnamefont{Brandes}},\ }%
  \bibfield{journal}{%
  {\bibinfo {journal} {Phys. Rev. B}}\ }%
  \textbf{\bibinfo {volume} {80}},\ \bibinfo {pages} {245107} (\bibinfo {year} {2009})%
  \bibAnnoteFile{NoStop}{Gernot2009}
%31
\bibitem{Livermore1996}%
  \BibitemOpen
  \bibfield{author}{%
  \bibinfo {author} {\bibfnamefont{C.}~\bibnamefont{Livermore}}, \bibinfo
  {author} {\bibfnamefont{C.~H.}\ \bibnamefont{Crouch}}, \bibinfo {author}
  {\bibfnamefont{R.~M.}\ \bibnamefont{Westervelt}}, \bibinfo {author}
  {\bibfnamefont{K.~L.}\ \bibnamefont{Campman}},\ and\ \bibinfo {author}
  {\bibfnamefont{A.~C.}\ \bibnamefont{Gossard}},\ }%
  \bibfield{journal}{%
  {\bibinfo {journal} {Science}}\ }%
  \textbf{\bibinfo {volume} {274}},\ \bibinfo {pages} {1332} (\bibinfo {year}
  {1996})%
  \bibAnnoteFile{NoStop}{Livermore1996}
%32
\bibitem{Cottet2004}%
  \BibitemOpen
  \bibfield{author}{%
  \bibinfo {author} {\bibfnamefont{A.}~\bibnamefont{Cottet}}, \bibinfo {author}
  {\bibfnamefont{W.}~\bibnamefont{Belzig}},\ and\ \bibinfo {author}
  {\bibfnamefont{C.}~\bibnamefont{Bruder}},\ }%
  \bibfield{journal}{%
  {\bibinfo {journal} {Phys. Rev. B}}\ }%
  \textbf{\bibinfo {volume} {70}},\ \bibinfo {pages} {115315} (\bibinfo {year} {2004})%
  \bibAnnoteFile{NoStop}{Cottet2004}
%33
\bibitem{Gernot2010}%
  \BibitemOpen
  \bibfield{author}{%
  \bibinfo {author} {\bibfnamefont{G.}~\bibnamefont{Schaller}}, \bibinfo
  {author} {\bibfnamefont{G.}~\bibnamefont{Kie\ss{}lich}},\ and\ \bibinfo
  {author} {\bibfnamefont{T.}~\bibnamefont{Brandes}},\ }%
  \bibfield{journal}{%
  {\bibinfo {journal} {Phys. Rev. B}}\ }%
  \textbf{\bibinfo {volume} {82}},\ \bibinfo {pages} {041303} (\bibinfo {year} {2010})%
  \bibAnnoteFile{NoStop}{Gernot2010}
%34
\bibitem{Jordan2004}%
  \BibitemOpen
  \bibfield{author}{%
  \bibinfo {author} {\bibfnamefont{A.~N.}\ \bibnamefont{Jordan}}\ and\ \bibinfo
  {author} {\bibfnamefont{M.}~\bibnamefont{B\"uttiker}},\ }%
  \bibfield{journal}{%
  {\bibinfo {journal} {Phys. Rev. Lett.}}\
  }%
  \textbf{\bibinfo {volume} {92}},\ \bibinfo {pages} {247901} (\bibinfo {year} {2004})%
  \bibAnnoteFile{NoStop}{Jordan2004}
%35
\bibitem{Wotters1998}%
  \BibitemOpen
  \bibfield{author}{%
  \bibinfo {author} {\bibfnamefont{W.~K.}\ \bibnamefont{Wootters}},\ }%
  \bibfield{journal}{%
  {\bibinfo {journal} {Phys. Rev. Lett.}}\ }%
  \textbf{\bibinfo {volume} {80}},\ \bibinfo {pages} {2245} (\bibinfo {year} {1998})%
  \bibAnnoteFile{NoStop}{Wotters1998}
%36
\bibitem{Gernot2011}%
  \BibitemOpen
  \bibfield{author}{%
  \bibinfo {author} {\bibfnamefont{G.}~\bibnamefont{Schaller}},\ }%
  \bibfield{journal}{%
  {\bibinfo {journal} {Phys. Rev. E}}\ }%
  \textbf{\bibinfo {volume} {83}},\ \bibinfo {pages} {031111} (\bibinfo {year} {2011})%
  \bibAnnoteFile{NoStop}{Gernot2011}
%
\end{thebibliography}
%
%
%
%
%
\end{document}